\documentclass[article]{jfm}
\usepackage{graphicx,colortbl,subcaption} 
\usepackage[normalem]{ulem} 
\usepackage{amsmath,amssymb}
\usepackage{pdfpages}
\usepackage{natbib}
\usepackage{booktabs}
\usepackage{gensymb}
\usepackage{tikz}

\usepackage[toc,page]{appendix}
\usepackage{newtxtext}
\usepackage{newtxmath}
\usepackage{hyperref}
\usepackage{romannum}
\usepackage{arydshln}
\usepackage{comment}
\usepackage[table]{xcolor}
\hypersetup{
    urlcolor   = blue,
    citecolor  = black,
}

\newcommand{\RomanNumeralCaps}[1]
\linenumbers

\title{Long-lived versus nocturnal stable atmospheric boundary layers: DNS characterisation, similarity theory, and regime classification}

\author{K. Chand \aff{1}
  Cheng-Nian Xiao \aff{1}
 \and Inanc Senocak\aff{1}
 \corresp{\email{senocak@pitt.edu}}}

\affiliation{\aff{1}Department of Mechanical Engineering and Materials Science, University of Pittsburgh, Pittsburgh, Pennsylvania}

\begin{document}

\makeatletter
\newenvironment{nbytwosubequations}{%
  \refstepcounter{equation}%
  \protected@edef\theparentequation{\theequation}%
  \setcounter{parentequation}{\value{equation}}%
  \setcounter{equation}{0}%
  \def\theequation{%
    \theparentequation\alph{equation}%
    \addtocounter{equation}{1},\alph{equation}%
  }%
}{%
  \setcounter{equation}{\value{parentequation}}%
}
\makeatother

\pagenumbering{arabic} 

\maketitle

\begin{abstract}
The stable atmospheric boundary layer (SABL) is broadly classified into two subtypes: the nocturnal SABL, driven primarily by surface cooling, and the long-lived SABL, in which ambient stratification coexists with surface cooling and persists well beyond a single diurnal cycle. Despite its prevalence in polar regions and over open oceans, the long-lived SABL has received comparatively little attention, and a systematic direct numerical simulation (DNS) study distinguishing it from the nocturnal SABL has not previously been reported. Here we present DNS of long-lived SABLs at $Re_D=900$, contrasting weak and strong ambient stratification cases against a nocturnal SABL. A striking feature of the long-lived SABLs is the emergence of a multi-layered thermal structure — including an intermediate layer of reduced static stability capped by a buoyancy inversion — driven not by enhanced turbulent mixing, but by a stratification-induced reorganisation of the buoyancy field. Budget analyses identify a turbulent potential energy-centred transport mechanism absent in nocturnal SABLs, lending support to total-energy-based closure approaches. Assessment of an extended Monin–Obukhov similarity theory incorporating a composite length scale $L_*$ shows strong collapse of dimensionless gradients of velocity $\phi_m$ and potential temperature $\phi_h$, and a new similarity function for $\phi_h$ is proposed. Combining DNS with linear stability analysis, we construct a regime map delineating linearly stable, weakly stable, and very stable regimes within a two-parameter dimensionless space, demonstrating the inherently multi-parameter nature of long-lived SABLs and the limitations of single-parameter subgrid-scale parameterisations.

\end{abstract}



\section{Introduction}\label{sec:intro}
The atmospheric boundary layer (ABL) is the lowest portion of the troposphere directly influenced by interactions with the Earth's surface \citep{stull_introduction_1988}. The canonical framework for studying fundamental processes within the ABL is the Ekman boundary layer, originally formulated for ocean currents \citep{ekman_influence_1905}, in which wind direction veers with decreasing height toward the surface as a consequence of the balance among viscous forces, the imposed pressure gradient, and the Coriolis force. The simplicity and analytical tractability of the Ekman model make it an enduring tool for isolating the essential dynamics of the ABL, and it has served as the foundation for a broad class of theoretical, numerical, and observational investigations of boundary-layer turbulence.

The structure of the ABL over land surfaces varies diurnally in response to buoyancy-driven processes. During the daytime, surface heating generates convective turbulence that promotes vigorous vertical mixing, whereas after sunset, radiative surface cooling --- particularly under clear-sky conditions --- suppresses turbulence and leads to the formation of a stable ABL (SABL), which typically intensifies throughout the night \citep{andre1982, ha2003}. Beyond radiative surface cooling, several additional mechanisms can drive SABL development, including warm-air advection \citep{mulhearn_warm_air_1981, Bonin2015greatplains}.

Historically, the ABL has been classified as stable, neutral, or unstable according to the vertical gradient of virtual potential temperature or the sign of the surface heat flux, a framework largely reflecting the canonical mid-latitude, land-based diurnal cycle in which convective daytime boundary layers transition to nocturnal SABLs capped by a residual layer \citep{stull_introduction_1988}. Within this traditional view, the terms ``nocturnal'' and ``stable'' boundary layer were often used synonymously. This classification, however, is inadequate for environments in which the static stability of the free atmosphere --- referred to here as ambient stratification --- acts independently of, and in addition to, surface cooling. Recognising this limitation, \citet{zilitinkevich2000, Zilitinkevich_third_order_2002, zilitinkevich_similarity_2007} distinguished two SABL types: the nocturnal SABL, driven primarily by surface cooling and typically transient in nature, and the long-lived SABL, in which ambient stratification coexists with near-surface cooling and persists well beyond the timescale of a single diurnal cycle.

Long-lived SABLs are especially prevalent in polar regions and over open oceans, where  surface temperature variations are weak and residual layers are absent \citep{bornstein2023}. Under such conditions, stable stratification can persist for extended periods --- from several days to multiple months \citep{Vignon2017, Dice2023} --- a regime that lies well outside the scope of the classical nocturnal SABL framework. 

In stably stratified flows, shear-generated turbulence competes with the suppressing effects of buoyancy, giving rise to complex flow dynamics characterised by intermittency, turbulence collapse, and subsequent resurgence — phenomena that motivate regime-based classification frameworks. The nocturnal SABL is commonly classified into two regimes — \textit{weakly stable} and \textit{very stable} — based on qualitative differences in turbulence behaviour \citep{mahrt_stratified_1998}. More refined classifications have since been proposed to further discriminate stable boundary layer behaviour across a broader range of stratification conditions \citep{grachev_stable_2005, sorbjan_evaluation_2010}. The weakly stable regime typically occurs under cloudy conditions or strong winds, where sufficient wind shear sustains turbulence that is continuous in both space and time. In contrast, clear skies and weak winds favour the very stable regime, in which turbulence is weak and intermittent; with increasing stratification strength, turbulence may collapse partially or completely, leading to flow relaminarisation \citep{mahrt_stratified_1998, mahrt_stably_2014}. 

The development of quantitative measures to identify turbulence collapse or relaminarization has attracted considerable research attention. Several studies have highlighted the limitations of using a critical gradient Richardson number  ($Ri_g$) to characterise turbulence suppression, as turbulence can persist even when $Ri_g >> 1.0$\citep{galperin_critical_2007, grachev_critical_2013, canuto_stably_2008, sorbjan_microstructure_2008}. Using DNS, \citet{nieuwstadt_direct_2005} investigated relaminarization in stably stratified open-channel flows and proposed $h/L <1.25$—where $h$ is the channel height and $L$ the viscous Obukhov length —  as a criterion for sustained turbulence, with flow collapse occurring beyond this threshold. Subsequently, \citet{flores_analysis_2011} demonstrated that  $h/L$ depends on $\Rey$ and proposed that a threshold value of $L u_{\ast}/\nu \approx 100$ corresponds to complete relaminarization in stably stratified open-channel flows, where $u_{\ast}$ is the friction velocity and $\nu$ is the kinematic viscosity.  In stably stratified channel flows, \citet{garcia-villalba_turbulence_2011} characterised flow regimes as a function of the friction Reynolds number and friction Richardson number, identifying distinct turbulent, laminar unstable, and laminar stable states. Collectively, these studies underscore the need for robust quantitative metrics to clearly demarcate weakly and very stable regimes.

A formal regime classification for the long-lived SABL does not yet exist. Given the simultaneous presence of ambient stratification and surface cooling, one might intuitively expect the \textit{very stable} regime to dominate; however, empirical evidence does not support this assumption \textit{a priori}. Indeed, observational evidence from Antarctica suggests that the relationship between stratification strength and sky conditions is more nuanced: \citet{Dice2023} identified twenty boundary layer regimes over Antarctica and found that while moderate and strong stability are preferentially associated with clear-sky conditions, weaker stability regimes occur with comparable frequency under both clear and cloudy skies. \citet{Vignon2017} observed sharp regime transitions from the weakly stable to the very stable over the Antarctic Plateau. Consistent with this observational complexity, the present study demonstrates that both the \textit{weakly stable} and \textit{very stable}, regimes can arise within a long-lived SABL, and that these regimes can be systematically distinguished within a suitably defined dimensionless parameter space.

The nocturnal SABL has received considerable attention in the literature \citep{coleman_numerical_1990, coleman_direct_1992, coleman_numerical_1994, hutchins_evidence_2007, deusebio_numerical_2014, ansorge_global_2014, shah_direct_2014, ansorge_analyses_2016, cheng_logarithmic_2023, greene_coherent_2024}. The pioneering DNS of neutral, convective, and stably stratified Ekman layers by \citet{coleman_numerical_1990, coleman_direct_1992, coleman_numerical_1994}, conducted at $Re_D = 400$ based on the laminar Ekman layer depth, established a foundational picture of the nocturnal SABL. Their results revealed a two-layer structure comprising an inner, shear-dominated region near the surface and an outer region in which Coriolis and buoyancy forces exert a strong influence on turbulence structure, anisotropy, and the veering of the mean velocity with height. Building on this configuration, \citet{shah_direct_2014} extended the DNS to Reynolds numbers of $Re_D =$ $600$, and $900$, elucidating a turbulent transport mechanism in which buoyancy indirectly attenuates TKE by suppressing vertical velocity fluctuations, thereby reducing vertical momentum fluxes and diminishing TKE production rather than acting through direct buoyancy-induced destruction. They further identified the inherent challenges in modelling the vertical velocity variance budget, particularly due to the complexity of pressure redistribution terms, and demonstrated that buoyancy influences turbulent heat transfer in a manner analogous to its effect on momentum -- consistent with the earlier findings of \citet{coleman_direct_1992} -- a result reflected in the near-constancy of $Pr_t$ throughout the bulk flow. As demonstrated in the present study, however, this constancy does not extend to long-lived SABLs, where $Pr_t$ exhibits dependence on both height and stability strength.

Monin--Obukhov similarity theory (MOST) \citep{monin_basic_1954} underpins the parameterisation of surface fluxes of heat, moisture, and momentum in numerical weather prediction models and large-eddy simulation codes. MOST provides universal flux-profile relationships within the surface layer -- approximately the lowest 10\% of the ABL -- under stationary, horizontally homogeneous conditions over flat terrain. The local scaling hypothesis of \citet{nieuwstadt_turbulent_1984} extends this framework to the bulk of the SABL, where surface-layer scaling no longer strictly applies. The flux-profile relationships used in conjunction with MOST have been developed and refined through field campaigns conducted over flat, homogeneous, mid-latitude terrain \citep{businger_flux-profile_1971, dyer_review_1974, hogstrom_non-dimensional_1988}. It is generally acknowledged that weakly stable nocturnal SABLs over such terrain are reasonably well represented by this framework. DNS of nocturnal SABLs have confirmed that MOST reliably captures flux-profile relationships in the weakly stable regime, although the empirical constant appearing in these relationships varies across investigations \citep{shah_direct_2014, ansorge_global_2014, gohari_direct_2017, deusebio_numerical_2014}.

Unlike the nocturnal SABL, the long-lived SABL has received relatively little attention, in part because its recognition as a distinct stable boundary layer subtype is relatively new \citep{bornstein2023}, and in part because it predominantly occurs in polar regions or over open oceans --- environments less accessible to routine observation than the mid-latitude land surfaces where nocturnal SABLs are commonly studied. High-fidelity simulations of long-lived SABL have been limited. As a precursor to the present DNS investigation of long-lived SABL, \citet{xiao_impact_2022} performed DNS of open-channel flows driven subject to both constant ambient stratification and surface cooling, introducing an expanded dimensionless parameter space that simultaneously accounts for both stratification mechanisms. By systematically varying the strength and type of stratification, they demonstrated that the resulting dimensionless mean profiles deviate substantially from MOST predictions, and that the extended similarity theory of \citet{zilitinkevich2000}, which explicitly accounts for ambient stratification, provides improved agreement with their DNS of open-channel flow results for dimensionless gradient of velocity, but performed poorly for the gradient of temperature. 


 \citet{Esau2004_LES} performed large-eddy simulations (LES) of ABL for truly neutral, conventionally neutral, nocturnally stable, and long-lived stable regimes., with the results compiled into DATABASE64 only several years later \cite{esau2010arxiv}. This database was employed to assess and calibrate resistance-law correlations \cite{esau_zilitinkevich_2006_universal}, equilibrium ABL height parameterisations \cite{zilitinkevich_further_2007}, and extension of MOST \cite{zilitinkevich_similarity_2007}. DATABASE64 was further used in the development of turbulence closure models for stable conditions based on the total turbulent energy concept \cite{zilitinkevich_turbulence_2008, Mauritsen2007_total_energy}. While these comparisons yielded improved correlations, the fidelity of the underlying LES — particularly with respect to surface parameterisations — warrants scrutiny. Specifically, the LES code used to generate DATABASE64 prescribes a constant surface heat flux, parameterises the surface momentum flux without MOST stability corrections (i.e. assuming neutral stratification near the surface), and imposes ambient stratification solely through the initial temperature profile \cite{esau2010arxiv}. For LES to yield physically reliable predictions, surface fluxes of heat and momentum must be parameterised using a similarity framework consistent with the stability conditions under investigation. DATABASE64 does not satisfy this requirement, and conclusions drawn from studies relying on it should therefore be interpreted with caution. Nonetheless, these investigations established the important role of ambient stratification in governing the structure of conventionally neutral and long-lived stable ABLs.

To the best of our knowledge, the present work represents the first DNS study of long-lived SABLs to systematically characterise and distinguish their turbulent structure from that of nocturnal SABLs. The principal contributions of this work are threefold:
\begin{enumerate}
    \item We provide the first DNS-based characterisation of long-lived SABLs 
    at $Re_D=900$ under weak and strong ambient stratification, revealing key 
    distinctions from nocturnal SABLs in mean flow structure, turbulent 
    statistics, and energy budgets (\S\ref{sec:meanprofiles} and 
    \S\ref{sec:budget});

    \item We present the first direct assessment of the extended similarity 
    theory of \citet{zilitinkevich_similarity_2007}, calibrating its parameters 
    against DNS data to show it can represent dimensionless momentum and scalar 
    gradients for both long-lived and nocturnal SABLs 
    (\S\ref{sec:MOST});

    \item We propose a regime classification for long-lived SABLs based entirely 
    on dimensionless parameters capturing the dominant stratification mechanisms 
    (\S\ref{sec:regime}).
\end{enumerate}
\section{Technical Formulation}\label{sec:numerics}
\begin{figure}
    \centering
    \includegraphics[width=0.7\linewidth]{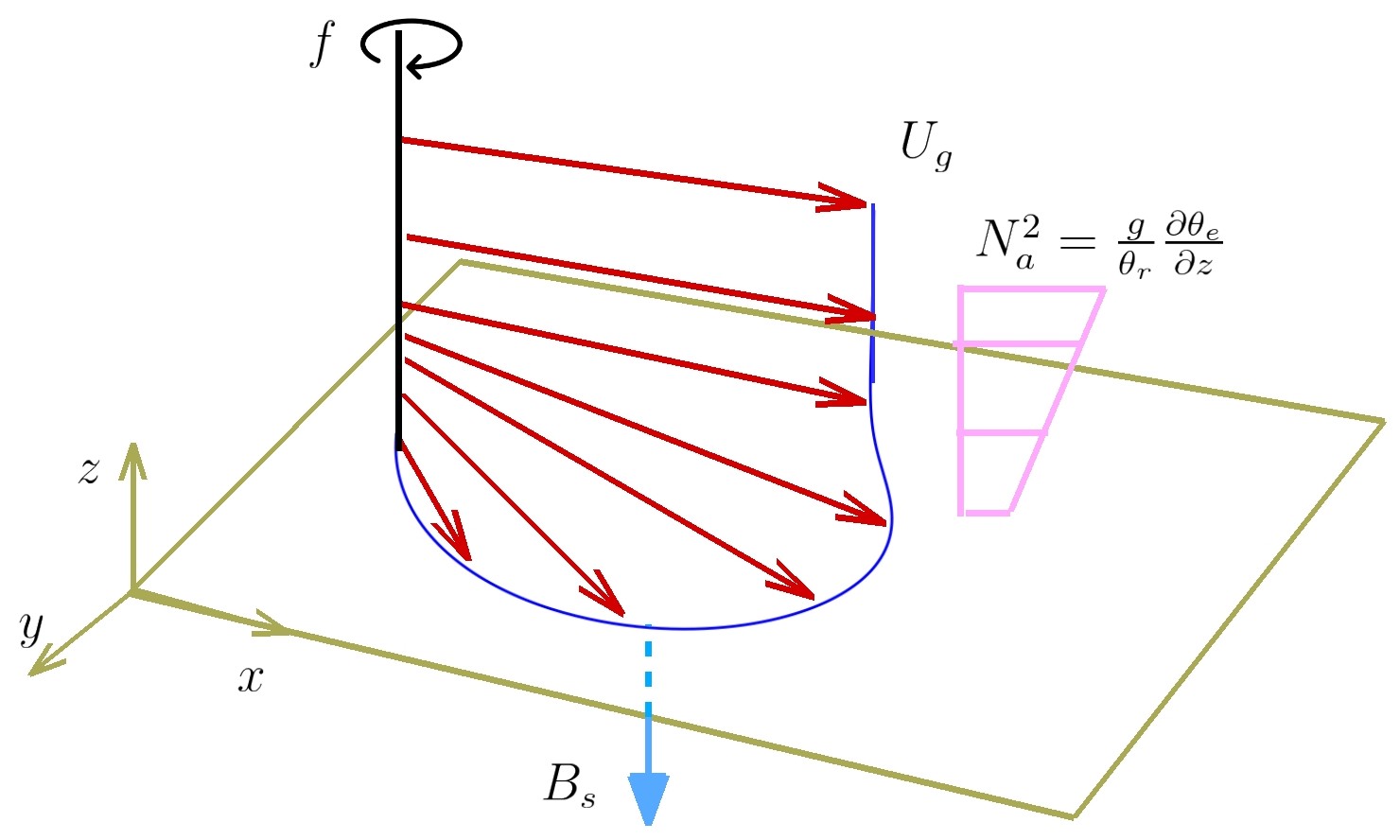}
    \caption{Illustration of the long-lived stable Ekman boundary layer with key parameters. $U_g$ is the geostrophic wind, $B_s$ is the imposed surface buoyancy flux, and $N_a^2$ is the strength of the ambient stratification that is imposed independent of $B_s$ in the equations.}
    \label{fig:schematic}
\end{figure}

\begin{table}
\footnotesize
\caption{Simulation parameters for $Re_D = 900$. From left to right: case identifier; dominant stratification mechanism (DSM); wind forcing parameter ($\Pi_w$); stratification perturbation parameter ($\Pi_s$); friction velocity $(u_*/U_g)$; friction Reynolds number $(\Rey_\tau)$; bulk Richardson number based on prescribed buoyancy flux $(Ri_b)$, and a posteriori calculated $(Ri^*_b)$ based on temperature difference ; number of grid points in the $x$–$z$ plane and in the $y$ direction; and the wall-unit grid spacings $\Delta_x^+$ and $\Delta_z^+$.} 
\label{tab:re900} 
\centering 
\begin{tabular}{@{}lllccccccccc@{}}
\toprule
    Cases & DSM & $\Pi_w$ & $\Pi_s$ & $u_*/U_g$ & $Re_\tau$& $Ri^*_b$& $N_x^2 \times N_z$ & $\Delta x^+(=\Delta y^+)$ & $\Delta z^+$ \\  \midrule  
Neutral \  & Neutral (N)          & --        & --  & 0.054    & 1208  & -- &   $320^2 \times 1664$ & 7.76 & 0.98 \\
Long-lived (S) \  & Strong $N_a$             & 25775        & 4     &0.0527    & 1121   & 0.72 &   $320^2 \times 1664$ & 7.48 & 0.89\\ 
Long-lived (W) \ & Weak $N_a$        & 47527        & 16  & 0.0528     & 1127  & 0.31 &   $320^2 \times 1664$ & 7.50 &  0.94 \\
Nocturnal \ & $B_s-$only     & --        & --   & 0.0528  & 1125  & 0.13 & $320^2 \times 1664$ & 7.49 &  0.95  \\ \hline
 \end{tabular}
\end{table}

The governing equations for a long-lived Ekman boundary layer flow are the conservation of mass,  momentum (under the Oberbeck-Boussinesq approximation), and energy
\begin{eqnarray}
\partial_{j} u_j&=&0 \label{eq:gov_eqn1}, \\
\partial_t u_i +  u_j \partial_j u_i \hspace{1pt} &=& -\partial_i p + \nu \partial^2_{jj} u_i + b \delta_{i3} + \epsilon_{ijk}(U_{g_j}-u_j)f_k, \\ \label{eq:gov_eqn2}
\partial_t b +  u_j \partial_j b \hspace{1pt} &=&  \beta \partial^2_{jj}b -{N_a^2} \delta_{_{j3}} u_{_3}, \label{eq:gov_eqn3}
\end{eqnarray}
where $t$ represents the time coordinate, $x_j$ ($j=1,2,3$) denotes the spatial coordinates, with ($j = 3$) indicating the vertical direction in which gravity acts, and $u_j$ = (u,v,w)  represents the velocity vector, $p$ is the pressure, $\delta_{i3}$ is the Kronecker delta, $\epsilon_{ijk}$ is the Levi-Civita symbol,  $\nu$ is the kinematic viscosity, $\beta$ is the thermal diffusivity. Buoyancy, $b=g\Phi$, is defined as the scaled potential temperature perturbation (deviation) with $\Phi=(\theta-\theta_e)/\theta_r$, where $\theta_e$ and $\theta_r$ are the environmental (ambient) and reference potential temperatures, respectively. Note that $\Phi=b/g$ is the dimensionless buoyancy and will be referred to as buoyancy hereinafter. Brunt-V{$\ddot{a}$}is{$\ddot{a}$}l{$\ddot{a}$} or buoyancy frequency is defined based on the profile of the ambient potential temperature as $N_a=\sqrt{\frac{g}{\theta_r}\tfrac{\partial \theta_e}{\partial z }}$ . $U_g$ is the geostrophic wind that is aligned with the x-direction, and $f$ is the Coriolis parameter. We invoke the $f-$plane approximation that corresponds to rotation of flow only along the vertical direction.
Note that the current formulation accounts for the ambient stratification---whose strength is quantified explicitly by $N_a$ in Eq. \ref{eq:gov_eqn3}. This framework enables two independent stratification mechanisms (surface cooling and ambient stratification) to act on the flow field simultaneously. This framework has also been adopted in stratified slope flows \cite{fedorovich_structure_2009, xiao_stability_2019}.

Figure \ref{fig:schematic} shows a schematic of a long-lived stable Ekman layer flow with. Periodic boundary conditions are imposed on the lateral boundaries (\(x\) and \(y\) directions), while  no-slip and symmetry boundary conditions are applied at the bottom (surface) and top domain boundaries, respectively. A constant, negative buoyancy flux $B_s$ is imposed at the surface, while adiabatic conditions are used at the top boundary. Computational domain sizes ($L_x \times L_y \times  L_z$) adopted in the present study are listed as a multiple of the laminar Ekman boundary layer depth \(D=\sqrt{2\nu/f}\), as shown in table \ref{tab:re900}.   

\subsection{Dimensionless Control Parameters}\label{sec:numerics_nondim_numbers} For a nocturnal stable atmospheric boundary layer (SABL), the Reynolds ($\Rey$), Richardson ($Ri$), Rossby ($Ro$), and Prandtl ($Pr$) numbers constitute the primary dimensionless control parameters. In contrast, for a long‑lived stable Ekman layer, the presence of an imposed background stratification—distinct from the surface‑generated cooling—introduces an additional nondimensional parameter that must be accounted for.

Application of the Buckingham–$\pi$ theorem to the governing equations for a long‑lived stable Ekman layer with a uniform surface cooling flux (i.e. equations \ref{eq:gov_eqn1}–\ref{eq:gov_eqn3}) yields the following set of dimensionless parameters:
\begin{equation}
    \Pi_s=\frac{|B_s|}{\beta {N_a}^2}=\frac{G_w}{{N_a}^2}, \hspace{10pt}\Pi_w=\frac{U_g^2}{\nu {N_a}}, \hspace{10pt}\Pi_f=\frac{f}{N_a}, \hspace{10pt}\Pran=\frac{\nu}{\beta},\label{eq:dim_param}
\end{equation}
where $G_w = \left( \frac{\partial b}{\partial z} \right)_{z=0}$ is the imposed buoyancy gradient in the surface (wall) normal direction. 
The nondimensional parameters $\Pi_s$ and $\Pi_w$ were first introduced in the context of canonical Prandtl slope flows \citeauthor{xiao_stability_2019}. The stratification‑perturbation parameter $\Pi_s$ represents the relative magnitude of the surface thermal forcing compared with the imposed ambient stratification \citep{xiao_stability_2019}. The wind‑forcing parameter $\Pi_w$ measures the ratio of kinetic‑energy input from the geostrophic (or ambient) wind to the combined damping arising from viscosity and background stratification \citep{xiao_linear_2020}. The Rossby‑radius parameter $\Pi_f$ characterizes the ratio of the Coriolis time scale to the buoyancy time scale associated with the ambient stratification. Given a characteristic height scale, $H$, the Rossby radius of deformation can be expressed as $L_R = H ~ \Pi^{-1}_f$. 

These dimensionless parameters may be further reorganized to reveal their relationship to the conventional nondimensional groups through appropriate internal scales, as follows:
\begin{equation}
    \Rey_D=\frac{U_g D}{\nu}=\sqrt{\frac{2\Pi_w}{\Pi_f}}, \hspace{5pt} Ri_{b,D} =\frac{2(1+\Pi_s)}{\Pi_w \Pi_f}, \hspace{5pt} Ri_{s,D} =\frac{|B_s| D^2}{\beta U_g^2}=\frac{2\Pi_s}{\Pi_w \Pi_f}=\frac{\Pi_s Re^2_D}{\Pi^2_w}, \label{eq:conv_param_old}
\end{equation}
where $Ri_{b,D}$ and $Ri_{s,D}$ denote the bulk and surface Richardson numbers, respectively, both defined using the laminar Ekman‑layer depth $D=\sqrt{2\nu/f}$. 

As evident from equation~\ref{eq:conv_param_old}, a parameter space based solely on $Re_D$ and $Ri_{b,D}$ (or $Ri_{s,D}$) is insufficient to characterize a long‑lived SABL, because each of these quantities depends implicitly on the expanded nondimensional parameter set defined in equation~\ref{eq:dim_param}. The parameters in equation~\ref{eq:dim_param} isolate the principal mechanisms governing the flow --- surface cooling, ambient stratification, rotation, and geostrophic wind forcing --- offering a physically transparent framework for interpreting the dynamics of long-lived, stably stratified Ekman layers.

To facilitate comparison with previous studies of the SABL \citep{coleman_direct_1992, shah_direct_2014}, the neutral turbulent Ekman layer length scale ($\delta_{t,N}=u_{\ast,N} f^{-1}$) is used as the characteristics length scale in the following dimensionless parameters
\begin{equation}
    \Rey_\tau=\frac{u_* \delta_{t,N}}{\nu}, \hspace{5pt} Ri_b =\frac{N_a^2+ G_w}{U_g^2}\delta_{t,N}^2, \hspace{5pt} Ri^*_b =\frac{g (\theta_s - \theta_\infty) \delta_{t,N}}{U_g^2}, \hspace{5pt} Ri_s =\frac{G_w}{U_g^2}\delta_{t,N}^2,\label{eq:conv_param}
\end{equation}
where $u_\ast$ is the friction velocity. $Ri_b$ is bulk Richardson number computed based on the ambient stratification and surface buoyancy flux. As a result, $Ri_b$ are larger than the values reported in literature. To address this, $Ri^*_b$ is computed a posteriori based on surface temperature $\theta_s$ and the temperature $\theta_{\infty}$ at height $\delta_{t,N}$, which facilitates direct comparison with the nocturnal SABL simulations of \citet{shah_direct_2014}. 

The principal distinction between our long-lived SABL setup and the nocturnal SABL of \citeauthor{shah_direct_2014} lies in the imposed ambient stratification and surface heat flux boundary condition, in contrast to the constant surface temperature used in their work and the absence of imposed ambient stratification.

\subsection{Numerical Method}
The governing equations \eqref{eq:gov_eqn1}-\eqref{eq:gov_eqn3} are solved using a spectral/hp‑element framework based on a continuous Galerkin–Fourier formulation, implemented in the open‑source package \texttt{Nektar++} (version 5.3.0) \citep{cantwell_nektar_2015, moxey_nektar_2020}. A uniform mesh of quadrilateral elements is constructed in the $x$–$z$ plane, with gravity aligned with the $z$‑axis. This two‑dimensional mesh is extruded in the spanwise direction ($y$) using Fourier spectral modes, employing the same resolution as in the $x$‑direction (see table \ref{tab:re900}). The spectral/hp‑element discretization uses a polynomial order of $p=9$, with the first grid point located at $z^+ \lesssim  1$ to ensure adequate near‑wall resolution (table \ref{tab:re900}). For the neutrally stratified case, this mesh yields $\Delta x^+ = \Delta y^+ = 7.76$ and $\Delta z^+ = 0.98$. Time advancement is performed using a second‑order implicit–explicit (IMEX) scheme, with numerical stability enforced by maintaining a Courant–Friedrichs–Lewy number below 0.3.
The adequacy of the spatial and temporal resolution was assessed by simulating a neutrally stratified Ekman layer and comparing the results with those of \citet{shah_direct_2014} (section \ref{sec:validation}). The validated computational setup was subsequently employed for all stratified cases listed in table \ref{tab:re900}.

To establish the initial conditions for all stratified simulations, a neutrally stratified baseline case was first conducted  over a minimum duration of six inertial cycles, where one inertial cycle is defined as $2\pi/f$. This integration period was sufficient to allow the flow to evolve from the prescribed initial state--comprising a laminar Ekman layer solution superimposed with random perturbations--into a fully developed turbulent regime, attaining a statistically stationary state. The resulting flow field was subsequently employed as the initial condition for all stratified cases, each of which was initialized using the following buoyancy profile:
\begin{equation}\label{Eq:buoy_initprofile}
    b(z) = \frac{a_*}{2} {\bigg{(}\frac{-\pi}{\log(0.01)}}\bigg{)}^{1/2} G_w \hspace{2pt}\text{erf} \bigg{(} \frac{z/a_*}{(-\log(0.01))^{-1/2}}\bigg{)},
\end{equation}
where $a* = 10D$ represents the vertical distance beyond which the buoyancy gradient becomes negligible. Each stratified simulation is further advanced for half an inertial cycle to attain a new statistically stationary state, after which data sampling is conducted over an additional half inertial cycle. 
Note that Eq. \ref{Eq:buoy_initprofile} is a modified version of the initial temperature profile used in \citet{coleman_direct_1992}, who used an initial lapse rate in place of $G_w$.

The \texttt{Nektar++} solver was parallelized using the Message Passing Interface (MPI) framework. All simulations were performed on a high-performance computing system equipped with AMD EPYC 7763 processors. A typical stratified case with a computational domain of size $50D \times 50D \times 33D$ required approximately eight days of wall-clock time when executed on 1920 processors.

\subsection{Validation}\label{sec:validation}
\begin{figure}
    \includegraphics[width=\linewidth]{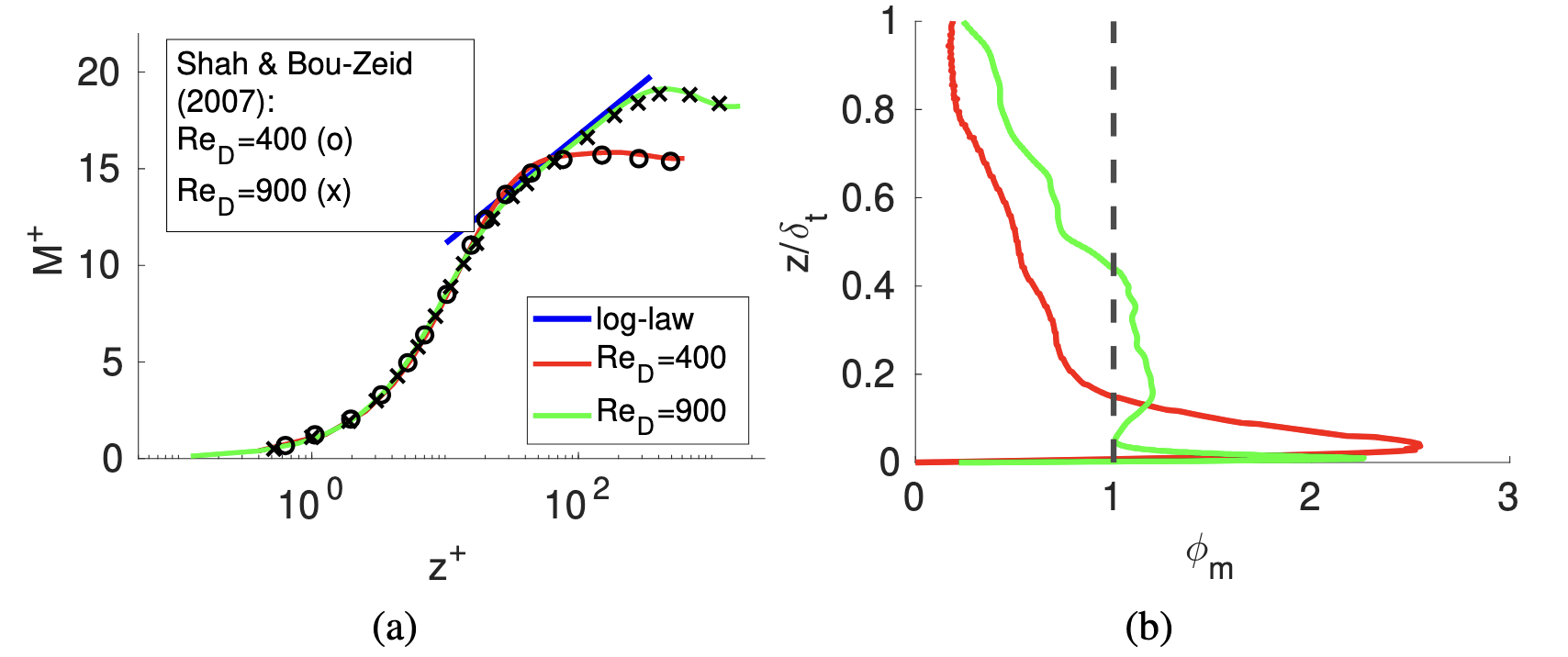}
\caption{Comparison of (a) the dimensionless velocity, $M^+ - z^+$, and (b) the dimensionless velocity gradient, $\phi_m$, under neutrally stratified conditions at $Re_D = 400$ and $900$, with the DNS results of \citet{shah_direct_2014}. The logarithmic law in panel (a) corresponds to $M^+ = \kappa^{-1}\ln(z^+) + B$, with $\kappa = 0.41$ and $B = 5.5$. The dashed line in panel (b) denotes $\phi_m = 1$.}
    \label{fig:validation_mplus}
\end{figure}

\begin{figure}
        \centering
        \includegraphics[width=0.48\textwidth]{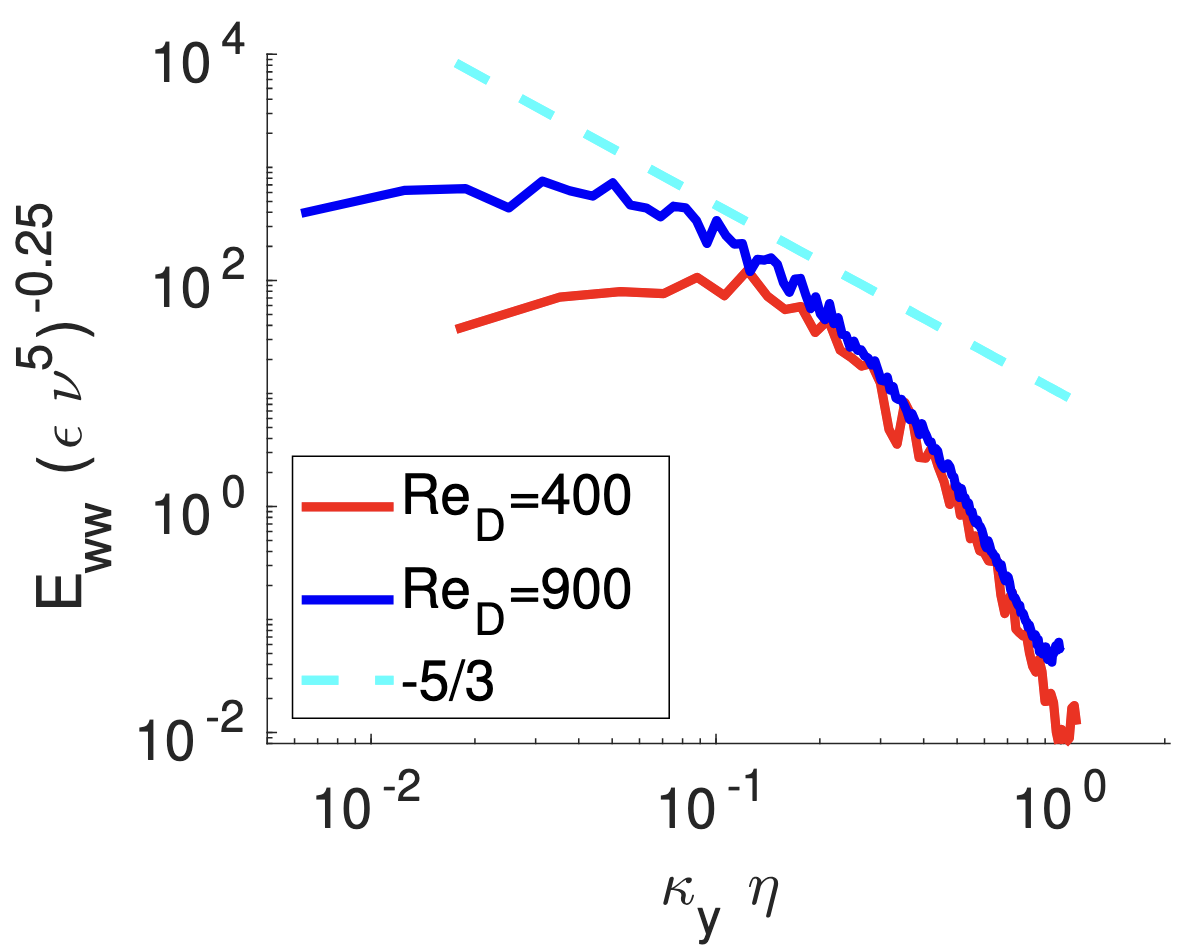}  
\caption{Dimensionless spectra of the vertical velocity, $E_{ww}$, at $z^+ = 32$ under neutrally stratified conditions for $Re_D = 400$ and $900$. Turbulence dissipation rate, $\epsilon$ and the Kolmogorov length scale are both evaluated at $z^+ = 32$.}
    \label{fig:validation_spectra}
\end{figure}

The \texttt{Nektar++} solver has been extensively validated across a broad range of flow configurations \citep{cantwell_nektar_2015, moxey_nektar_2020}, including applications to environmental and stratified shear flows \citep{xiao_stability_2019, xiao_impact_2022}. In addition, we verified both the solver and the adopted numerical resolution by comparing simulations of a neutrally stratified turbulent Ekman layer with the results of \citet{shah_direct_2014}.

Figure~\ref{fig:validation_mplus}(a) shows the variation of the dimensionless velocity $M^{+}$ with the dimensionless wall‑normal coordinate $z^{+}$, while figure~\ref{fig:validation_mplus}(b) presents the corresponding profiles of the dimensionless velocity gradient $\phi_m$ for $Re \in \{400, 900\}$ for neutrally stratified conditions. These quantities and coordinates are defined as


\begin{align}
    M^+ &= \frac{\sqrt{\langle u \rangle^2 + \langle v \rangle^2}}{u_*},\\
    u_*^2 &= \nu \left[ \left( \frac{\partial \langle u \rangle}{\partial z} \right)^2_{z=0} + \left( \frac{\partial \langle v \rangle}{\partial z} \right)^2_{z=0} \right]^{1/2},\\
    z^+ &= \frac{z u_*}{\nu}, \quad \delta_t = \frac{u_*}{f}, \label{eq:yplus} \\ 
    \phi_m &= \frac{z \kappa}{u_*} \left[ \left( \frac{\partial \langle u \rangle}{\partial z} \right)^2 + \left( \frac{\partial \langle v \rangle}{\partial z} \right)^2 \right]^{1/2}, \label{eq:phi_m}
\end{align}
where $\langle \cdot \rangle$ denotes averaging over the horizontal plane and time. Here, $u_*$ is the friction velocity, $\delta_t$ is the turbulent Ekman layer depth. The von Kármán constant is taken as $\kappa = 0.40$.

Dimensionless gradient of the potential temperature, $\Phi_h$, used later in section \ref{sec:MOST}, is computed as:
\begin{align}
    \phi_h &= \frac{z \kappa}{b_*} \left[ \frac{\partial \langle b \rangle}{\partial z}  + N_a^2 \right], \label{eq:phi_h}
\end{align}
where friction buoyancy is defined as $b_* = \beta G_w u^{-1}_*$.

As shown in figure~\ref{fig:validation_mplus}(a), the simulation results for both $Re_D$ cases closely reproduce the DNS data of \citet{shah_direct_2014}. For $Re_D = 900$, the mean velocity profile follows the logarithmic law over a wider range of $z^{+}$. The behaviour of the non‑dimensional shear function $\phi_m$ is further examined in figure~\ref{fig:validation_mplus}(b).
At the lower Reynolds number ($Re_D = 400$), the extent of the logarithmic region is limited, and agreement with MOST--specifically, $\phi_m = 1$ throughout the vertical domain--is therefore not expected. In contrast, the $Re_D = 900$ case exhibits good agreement with MOST within the range $0.2 < z/\delta_t < 0.4$.

The energy spectra of vertical velocity for both $Re_D$ cases are shown in figure~\ref{fig:validation_spectra}. Both the energy and wavenumber are normalized using the kinematic viscosity, viscous dissipation rate $(\epsilon)$, and the Kolmogorov length scale $(\eta)$, which are defined as follows:
\begin{equation}
    \epsilon = 2 \nu \langle \partial_j u_i^\prime \partial_j u_i^\prime \rangle; \hspace{2pt} \text{and} \hspace{2pt} \eta = \left( \frac{\nu^3}{\epsilon} \right)^{1/4}.
\end{equation} 

For $Re_D = 900$, the vertical‑velocity spectra exhibit a sufficiently extended inertial subrange that follows the theoretical $-5/3$ slope, indicating that this Reynolds number provides a well‑resolved representation of the turbulence. Consequently, $Re_D = 900$ is adopted for all simulations used to examine the turbulent characteristics and energy budget of long‑lived SABLs. Owing to its lower computational cost, $Re_D = 400$ is employed only to illustrate the multi‑parameter dependence of long‑lived SABLs for the development of a regime map.

\section{Results}
\subsection{Turbulent statistics}\label{sec:meanprofiles}
A central objective of this section is to establish how the presence of ambient stratification with constant strength fundamentally alters the turbulent structure of long-lived SABLs relative to the nocturnal SABL. To this end, four simulations are conducted at $Re_D = 900$: a neutrally stratified reference case, a nocturnal SABL driven by a constant negative surface heat flux, and two long-lived SABL cases with different ambient stratification in addition to negative surface heat flux. The governing parameters for all cases are provided in table~\ref{tab:re900}.

\begin{figure}
    \centering  
    \includegraphics[width=0.9\linewidth]{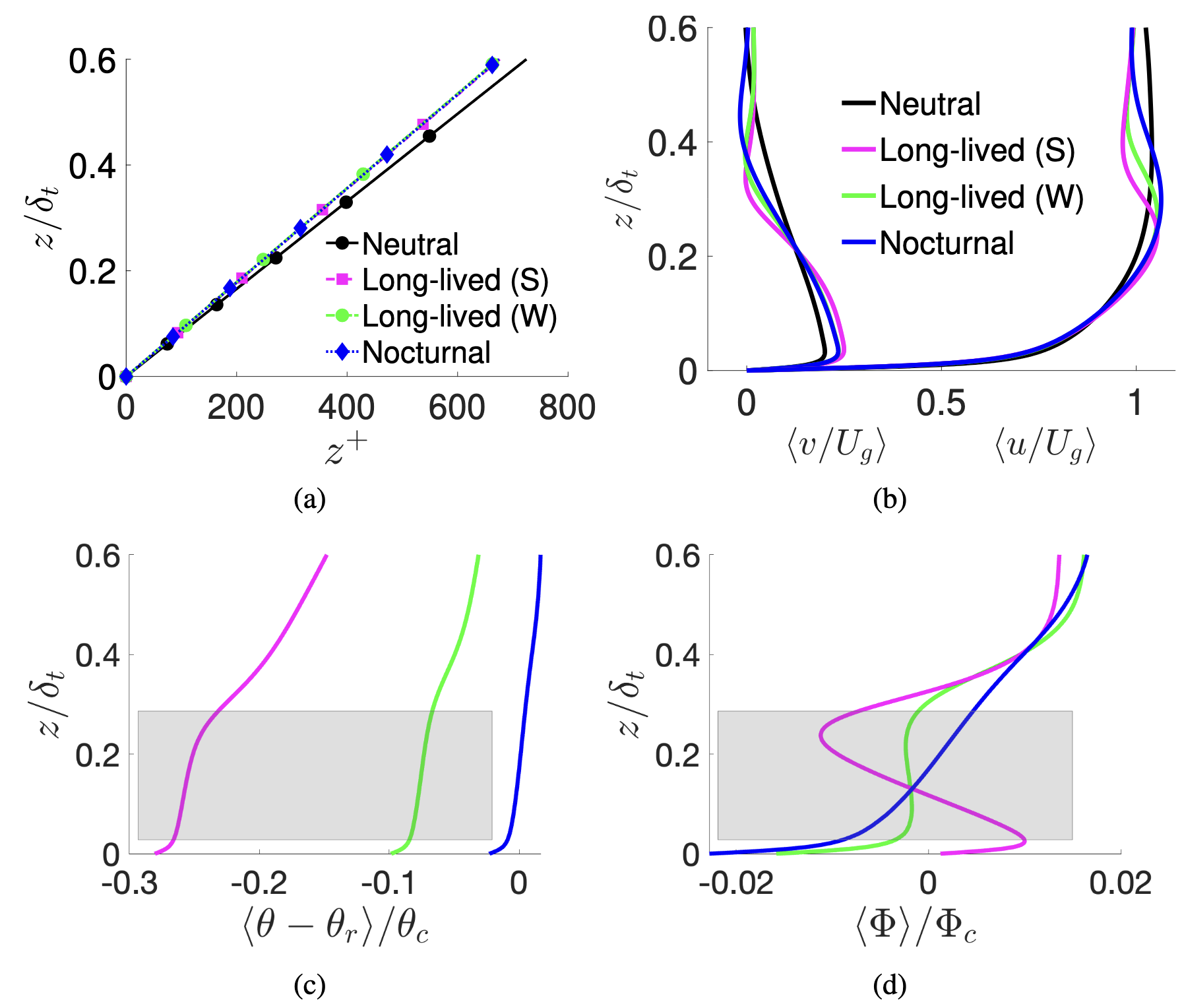}
    \caption{(a) Relationship between $z/\delta_t$ and $z^+$. Vertical profiles of mean (b) horizontal velocities ($\langle u \rangle$ and $\langle v \rangle$) and (c) potential temperature deviation $\langle \theta -\theta_r \rangle$ from reference temperature $\theta_r$, and (d) dimensionless buoyancy $\langle \Phi \rangle$. Long‑lived cases with strong (S) and weak (W) ambient stratification are indicated as defined in table~\ref{tab:re900}. The shaded region in panels (c) and (d) highlights the intermediate layer that develops in long‑lived SABLs and is absent in the nocturnal SABL.}
    \label{fig:meanprofiles_U_b}
\end{figure}

Figure~\ref{fig:meanprofiles_U_b}a shows the relationship between the outer-scaled height $z/\delta_t$ and the wall-unit coordinate $z^+ = z u_*/\nu$ for the three stratified cases. The two coordinates are related linearly through $z^+ = Re_\tau\,(z/\delta_t)$, and since the friction Reynolds numbers differ only marginally across the stratified cases ($Re_\tau \approx 1121$--$1127$; table~\ref{tab:re900}), the curves nearly collapse onto a single line. This provides a direct conversion between the inner- and outer-scaled coordinates used elsewhere in the analysis. 

Figure~\ref{fig:meanprofiles_U_b}b shows vertical profiles of the streamwise and spanwise components of the mean velocity for all four cases. For both the streamwise $\langle u \rangle$ and spanwise $\langle v \rangle$ velocity components, increasing ambient stratification promotes the development of a low-level jet, evident from the characteristic nose-shaped peak, and the streamwise velocity maximum shifts downward with stronger $N_a$.  

The corresponding mean potential temperature $\langle \theta \rangle$ and dimensionless mean buoyancy $\langle \Phi \rangle$ profiles are presented in figure~\ref{fig:meanprofiles_U_b}(c-d). The potential temperature $\theta$ is related to the buoyancy $b$ through
\begin{equation}
    \theta = \frac{\theta_r}{g} \left( b-N_a^2 \left(L_z-z\right)\right)+\theta_r, \label{eq:b_to_theta}
\end{equation}
where $L_z$ is the height of computational domain.  
Since both surface buoyancy flux and ambient stratification are present, the profiles are normalised using the composite scales $\theta_c$ and $\Phi_c$, defined as
\begin{equation}
   \theta_c = \frac{\theta_r}{g} (N_a^2+G_w) \delta_t \hspace{2pt}, \qquad \hspace{2pt}   \Phi_c = \frac{N_a^2+G_w}{g}\delta_t. 
\end{equation}

The mean potential temperature profiles of the long-lived SABL cases, shown in Figure~\ref{fig:meanprofiles_U_b}c, exhibit a distinct three-layer thermal structure. Close to the wall ($z/\delta_t \lesssim 0.05$), surface cooling produces a strongly stable near-wall region. Above this lies an intermediate layer (shaded region in Figure~\ref{fig:meanprofiles_U_b}c), where the static stability is reduced relative to the near-wall region. This intermediate layer is capped by an elevated buoyancy inversion aloft, as evident from Figure~\ref{fig:meanprofiles_U_b}d. In contrast, the nocturnal SABL exhibits a simpler two-layer structure, in which potential temperature and buoyancy increase approximately linearly with height above the near-wall region. The emergence of this three-layer structure is therefore a defining characteristic of the long-lived SABL.

Because the temperature equation is formulated in terms of buoyancy (equation~\ref{eq:gov_eqn3}), with potential temperature recovered via equation~\ref{eq:b_to_theta}, the layered structure is more clearly revealed by the buoyancy profiles shown in Figure~\ref{fig:meanprofiles_U_b}d. These profiles exhibit an even more pronounced stratification: in the long-lived SABL cases, the buoyancy gradient becomes negative within the intermediate layer (shaded region), with its magnitude increasing as the ambient stratification strengthens. In contrast, the buoyancy gradient remains positive throughout the depth of the nocturnal SABL.

\begin{figure}
\centering
    \includegraphics[width=0.85\linewidth]{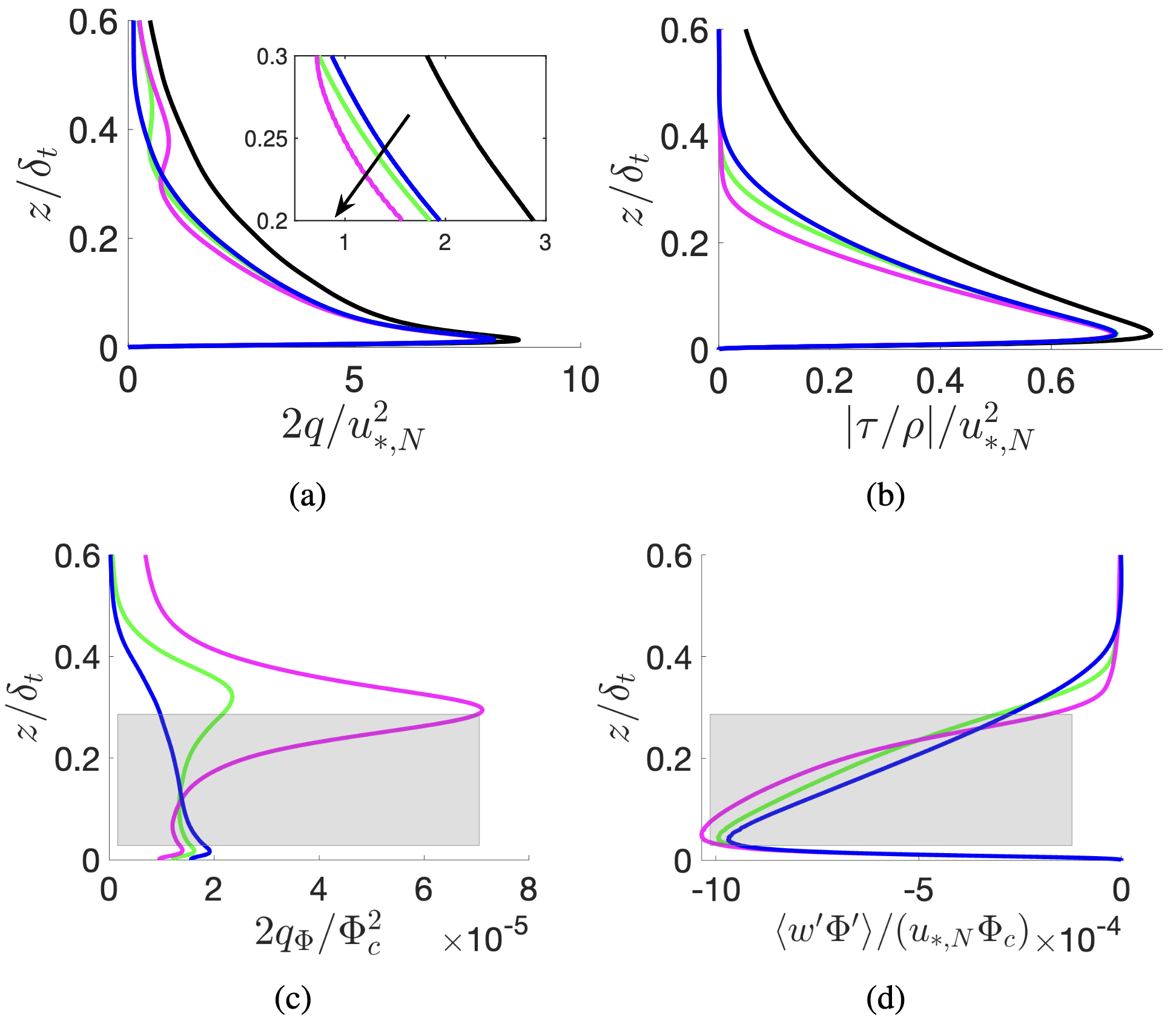}
    \caption{Vertical profiles of (a) TKE (b) Reynolds stress (c) TPE, and (d) turbulent buoyancy flux. The reference velocity and composite buoyancy scales are $u_{,N}$ and $\Phi_c$, respectively. The shaded regions in panels (c) and (d) highlight the intermediate layer that develops in long‑lived SABLs and is absent in the nocturnal SABL. See figure~\ref{fig:meanprofiles_U_b} for the legend.}
    \label{fig:meanprofiles_2ndorder}
\end{figure}

Figure~\ref{fig:meanprofiles_2ndorder} presents vertical profiles of the mean turbulent kinetic energy (TKE), $q = 0.5 \langle u_i^{\prime 2} \rangle$, and turbulent potential energy (TPE), $q_{\Phi} = 0.5 \langle \Phi^{\prime 2} \rangle$ --- a scaled measure of buoyancy variance --- together with the Reynolds stress, $\tau$, and the turbulent buoyancy flux, $\langle w' \Phi' \rangle$.

Figure~\ref{fig:meanprofiles_2ndorder}a shows that the TKE responds non-monotonically to the strength of ambient stratification. Despite the reduced static stability within the intermediate layer (identified in Figure~\ref{fig:meanprofiles_U_b}d), TKE levels in this region are slightly lower than those in the nocturnal case. In contrast, within the inversion layer above, where the buoyancy gradient is strongly positive, TKE levels exceed those of the nocturnal case. The Reynolds stress profiles, shown in Figure~\ref{fig:meanprofiles_2ndorder}b, exhibit a weakening trend with increasing ambient stratification.

Despite the moderate influence of ambient stratification on TKE and Reynolds stress profiles, its presence has a much stronger influence on the buoyancy statistics. As shown in figure~\ref{fig:meanprofiles_2ndorder}c, the TPE levels peak at the inversion layer with increasing strength of ambient stratification. These elevated TPE levels, however, rapidly diminish toward the levels of the nocturnal case in the intermediate layer. The turbulent buoyancy flux profiles in figure~\ref{fig:meanprofiles_2ndorder}d further corroborate this behavior. In particular, the downward turbulent buoyancy flux is markedly enhanced for the long‑lived SABL with the strongest ambient stratification.


Collectively, the profiles in figure~\ref{fig:meanprofiles_2ndorder} delineate the key structural differences between the long-lived and nocturnal SABLs. The defining signature of the long-lived SABL is a three-layer thermal structure comprising a strongly stable near-wall region, an intermediate layer of reduced static stability with a negative buoyancy gradient, and a capping inversion aloft. The implications of this structure on turbulent energy transport is further investigated in the next section.

\subsection{Turbulent kinetic and potential energy budgets}\label{sec:budget}
\begin{figure}
    \centering
    \includegraphics[width=\linewidth]{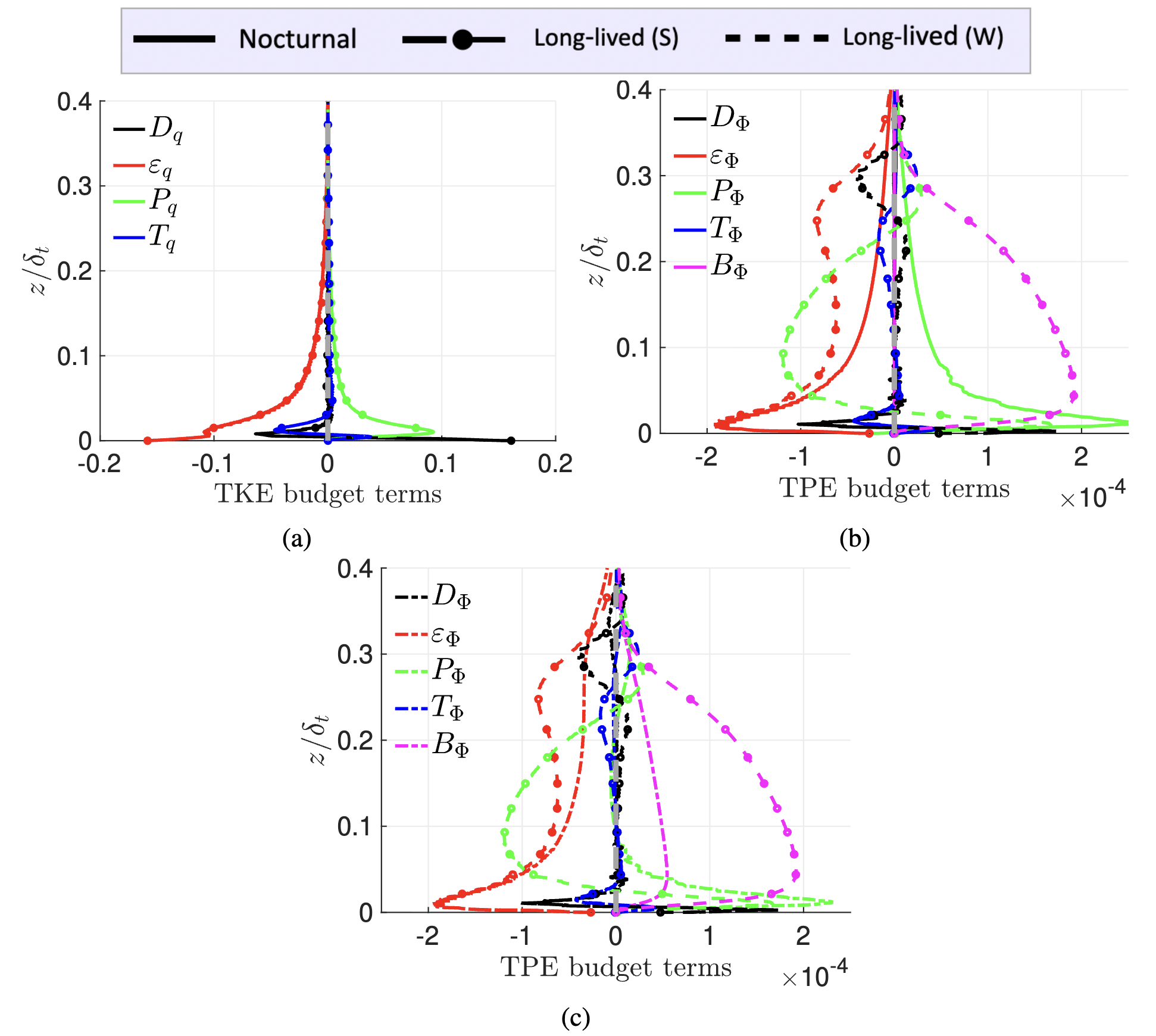}
    \caption{Comparison of the budget‑equation terms for (a) TKE, (b) TPE for the nocturnal and long‑lived (S) cases and (c) TPE for the long-lived (S) and (W) cases. The TKE budget includes diffusion ($D_{q}$), dissipation ($\epsilon_{q}$), shear production ($P_{q}$), and turbulent transport ($T_{q}$). Corresponding terms in the TPE budget use the same notation with the subscript $\Phi$, with an additional buoyancy‑production term ($B_\Phi$) arising from $N_a$. The grey vertical line denotes zero, separating regions of energy gain (positive values) and loss (negative values). TKE budget terms are normalized by $u_{*,N}^4/\nu$, while TPE terms are normalized by $(Pr \hspace{2pt} u_{*,N}^2/\nu) \Phi_c^2 $}
    \label{fig:budgetterms}
\end{figure}

The preceding analysis has shown that ambient stratification strongly influences thermal‐energy transport within a long‑lived SABL. In view of the three-layer thermal structure of long-lived SABLs, we now examine the budget equations for TKE and TPE. The full budget equations are given in Appendix~\ref{sec:appendix_budget}.  

Assuming horizontal homogeneity and the absence of subsidence, TKE and TPE budget terms are
\begin{nbytwosubequations}
\begin{eqnarray}
    B_{q} = g\langle w^\prime \Phi^\prime\rangle, &\qquad& B_{\Phi} = -N_a^2 g^{-1} \langle w^\prime \Phi^\prime\rangle, \\ 
    D_{q} = \nu \partial^2_{jj} \langle u^\prime_i u^\prime_i \rangle, &\qquad& D_{\Phi} = \beta \partial^2_{jj} \langle \Phi^{\prime~2} \rangle, \\ 
    \epsilon_{q} = -\nu \langle \partial_j u^\prime_i \partial_j u^\prime_i \rangle, &\qquad& 
    \epsilon_{\Phi} = -\beta \langle \partial_j \Phi^\prime \partial_j \Phi^\prime \rangle, \\ 
    P_{q} = -\langle u^\prime_i u^\prime_j \rangle \partial_j \langle u_i\rangle, &\qquad& P_{\Phi} = -\langle w^\prime \Phi^\prime \rangle \partial_z \langle \Phi\rangle, \\ 
    T_{q} = -0.5~\partial_j \langle u^\prime_j u^\prime_i u^\prime_i \rangle, &\qquad& T_{\Phi} = -0.5~\partial_j \langle u^\prime_j \Phi^\prime \Phi^\prime \rangle,
\end{eqnarray} \label{eq:budget_terms}
\end{nbytwosubequations}
\noindent where $B_{q}$, $D_{q}$, $\epsilon_{q}$, $P_{q}$, and $T_{q}$ denote buoyancy destruction (sink), viscous diffusion, dissipation, shear production (source), and transport of TKE, respectively. Corresponding terms in the TPE budget are denoted with the subscript $\Phi$. 

TKE and TPE budget equations are coupled through the turbulent buoyancy flux, $\langle w^\prime \Phi^\prime \rangle$, consistent with the view that total turbulent energy provides a more complete framework for turbulence closure models \citep{zilitinkevich_turbulence_2008}. 
In addition to $P_\Phi$, an extra buoyancy‑production term, $B_\Phi = -N_a^2 g^{-1} \langle w^\prime \Phi^\prime \rangle$, arises due to the imposed ambient stratification. This term is analogous to $P_\Phi$ but extracts energy from the background potential‑temperature profile associated with $N_a$, rather than from the mean buoyancy gradient. We refer to this contribution as buoyancy production due to $N_a$.

A direct comparison of TKE buoyancy production term $B_q$ with TPE budget terms $B_\Phi$ and $P_\Phi$ shows that ambient stratification $N_a$ and the mean buoyancy gradient $\partial_z \langle \Phi \rangle$ do not appear explicitly in the TKE budget, whereas both quantities modulate the effect of downward buoyancy flux in the TPE budget. Interestingly, $P_\Phi$ term can change from a source to a sink due to the sign reversal of the mean buoyancy gradient, as shown in figure \ref{fig:meanprofiles_U_b}d. This distinction explains the relatively modest response of the TKE and Reynolds stress profiles to ambient stratification, in contrast to the pronounced changes observed in thermal quantities for long-lived SABL cases.

\begin{figure}
    \centering
    \includegraphics[width=\linewidth]{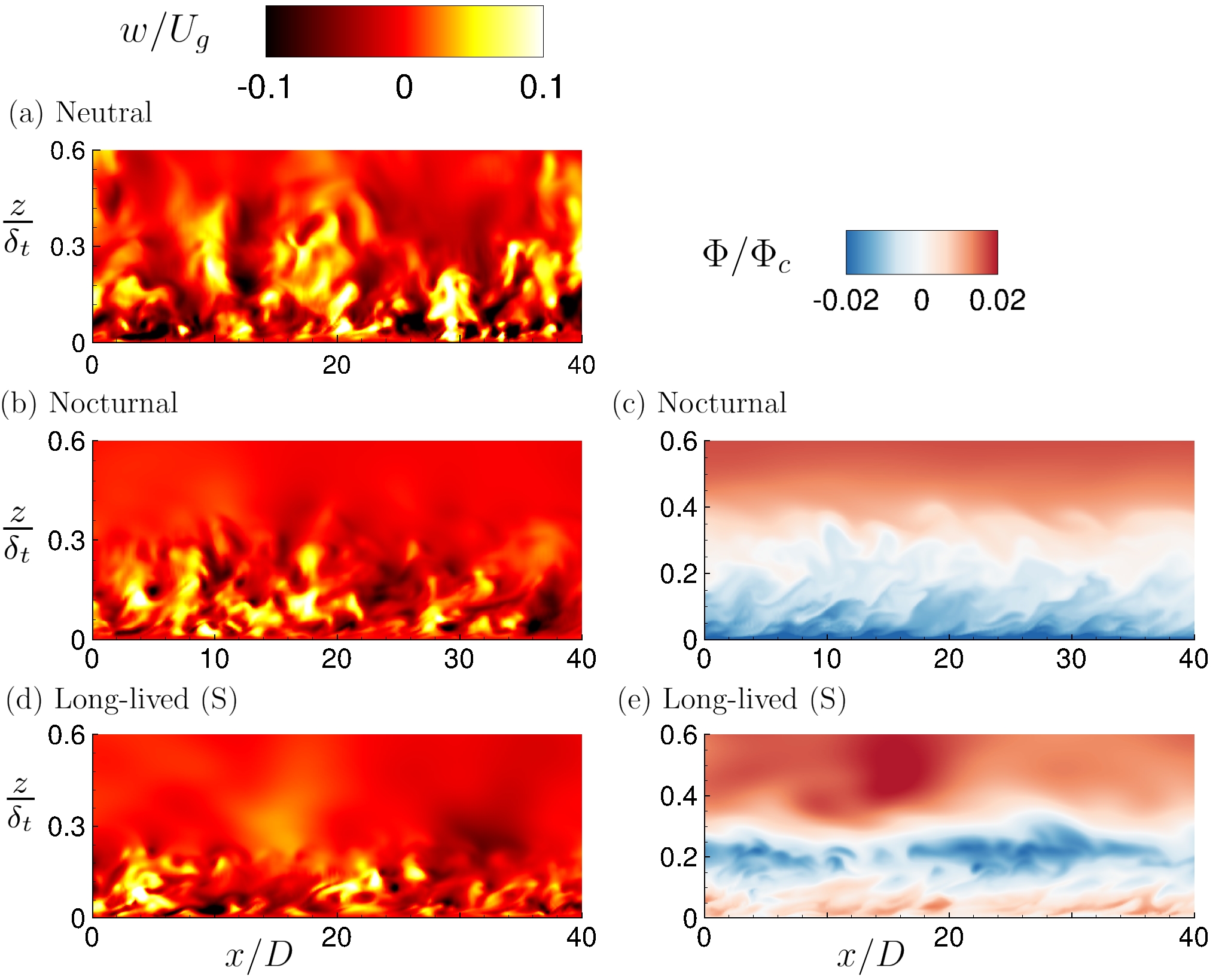}
    \caption{Contour visualizations of instantaneous vertical velocity (left) and buoyancy (right) for neutral, nocturnal, and long-lived SABL.}
    \label{fig:contourplots}
\end{figure}
Figure~\ref{fig:budgetterms}a contrasts the vertical profiles of the TKE budget terms for the nocturnal and long-lived (S) cases. Despite the differing stratification mechanisms, the TKE budget terms retain largely similar vertical structures across both cases. This insensitivity is consistent with the minimal influence of stratification mechanisms observed in the mean velocity profiles.  Moreover, in agreement with previous studies \citep{shah_direct_2014, gohari_direct_2017}, the primary balance in the TKE budget is maintained between production and dissipation.

Figure~\ref{fig:budgetterms}b compares the TPE budget profiles for the nocturnal SABL and long-lived (S) cases, revealing markedly different dynamics, particularly within the intermediate layer, $0.05 \lesssim z/\delta_t\lesssim 0.35$. Consistent with the three-layer thermal structure, the budget terms near the surface collapse for all cases, indicating the dominance of surface cooling in the near-wall region. With increasing height, however, the TPE budget becomes progressively influenced by the prevailing stratification mechanism.

In the nocturnal case, the TPE budget terms diminish rapidly aloft, reflecting a characteristic feature of nocturnal SABLs. In contrast, the magnitudes of the budget terms increase from the nocturnal to the long-lived (S) case. Notably, the long-lived (S) case exhibits an additional buoyancy production term $B_{\Phi}$, which is absent in the nocturnal SABL. The dominant contributions to the TPE budget are dissipation $\epsilon_{\Phi}$, shear production $P_{\Phi}$, and buoyancy production $B_{\Phi}$, as defined in Eq.~\ref{eq:budget_terms}. The combined production term ($P_{\Phi} + B_{\Phi}$) closely mirrors the dissipation profile $\epsilon_{\Phi}$ with opposite sign, indicating an approximate balance between total production and dissipation.

Interestingly, $P_\Phi$ in figure~\ref{fig:budgetterms}b undergoes a role reversal in the long-lived (S) case, acting as a source near the wall and as a sink aloft. This role reversal for $P_\Phi$ is much weaker in long-lived (W) case as seen in figure~\ref{fig:budgetterms}c, consistent with the buoyancy and turbulent buoyancy flux profiles shown earlier in figures~\ref{fig:meanprofiles_U_b}d and \ref{fig:meanprofiles_2ndorder}d. The influence of ambient stratification becomes more pronounced away from the surface, at $z/\delta_t \gtrsim 0.05$ (or $z^+ \gtrsim 33$; figure~\ref{fig:meanprofiles_U_b}a), where it gives rise to the formation of the intermediate layer. To maintain the energy balance in this region, $P_\Phi$ redistributes the enhanced TPE production by modulating the curvature of the mean buoyancy gradient, thereby sustaining the intermediate layer.

We further contrast the TPE budget terms between the long-lived (S) and (W) cases in figure \ref{fig:budgetterms}c. Compared to the long-lived (S), the distinct features (enhanced dissipation and role-switching of $P_\Phi$ at higher altitudes) of long-lived SABL is weaker in the (W) case. This explains the weakening of the buoyancy inversion in figure \ref{fig:meanprofiles_U_b}d. Also, the overall transport mechanism $T_{\Phi}$ in long-lived (W) exhibits the features of both the nocturnal and long-lived (S).


Figure~\ref{fig:contourplots} presents the vertical velocity and buoyancy contours for the nocturnal and the long-lived (S) cases along with vertical velocity field of the neutral case. From the neutral case to the long-lived (S) case (left column), vertical velocity fluctuations are suppressed, confining turbulent activity to the proximity of the surface. The buoyancy contours shown in figure~\ref{fig:contourplots}e reveal the three-layer structure of long-lived SABL with a buoyancy inversion forming at $z/\delta_t \approx 0.3$. This qualitative comparison shows that SABL height is much lower in a long-lived SABL. 

Our analysis of the TKE and TPE budgets indicates that turbulent transport in a long-lived SABL, shaped by ambient stratification, is considerably more complex than in a nocturnal SABL within the core of the boundary layer. In this regard, the TPE budget provides greater insight than the TKE budget. However, in the near-surface region ($z/\delta_t \lesssim 0.047$, or $z^+ \lesssim 52$; Figure~\ref{fig:meanprofiles_U_b}a), the transport mechanism in both types of SABL is governed primarily by surface cooling.

\subsection{Implications for Similarity Theory}\label{sec:MOST}
Monin--Obukhov similarity theory (MOST), together with the associated flux-profile relationships, forms the backbone of surface-flux parameterizations of momentum and scalars used in numerical weather prediction (NWP) and large-eddy simulation (LES) of the ABL. Given its practical importance, we examine the dimensionless velocity ($\phi_m$, see equation~\ref{eq:phi_m}) and potential-temperature ($\phi_h$, see equation~\ref{eq:phi_h}) gradients for the nocturnal and long-lived SABLs and assess them against an extended version of MOST by \citet{zilitinkevich_similarity_2007}, which is formulated to account for ambient stratification --- referred to in their work as the static stability of the free atmosphere --- and Coriolis forcing.

A cornerstone of the extended MOST by \citeauthor{zilitinkevich_similarity_2007} is the following composite length scale
\begin{align}
    \frac{1}{L_*} &= \left[\left(\frac{1}{L_o}\right)^{2} + \left(\frac{C_N}{L_N}\right)^{2} + \left(\frac{C_f}{L_f}\right)^{2}\right]^{1/2}, \label{eq:most_extension_c}
\end{align}
\begin{equation}
    L_o = \frac{-u_*^3}{\kappa B_s} \qquad L_N = \frac{u_*}{N_a}, \qquad L_f = \frac{u_*}{f}, \label{eq:most_ex_scales}
\end{equation}
where $L_o$ is the Obukhov length, $L_N$ and $L_f$ are length scales characterizing the effect of the ambient stratification and Earth's rotation, respectively. For the nocturnal SABL ($N_a = 0$), we drop out the $C_N/L_N$ term from the composite length scale, so the same formula remains applicable across all three cases considered here.

An important point to note is that \citeauthor{zilitinkevich_similarity_2007} originally proposed a height-dependent versions of the length scales given by \ref{eq:most_ex_scales}, consistent with the local scaling hypothesis \cite{nieuwstadt_turbulent_1984}, in which surface friction velocity $u_*$ and surface buoyancy flux $B_s$ are replaced by the local Reynolds stress $\sqrt{\tau(z)}$ and buoyancy flux $B(z)$. Owing to the relatively low Reynolds number of our DNS, we adopt the conventional surface-layer scaling assumption and use $u_*$ and $B_s$ as the characteristic velocity and buoyancy scales, respectively. 

\citeauthor{zilitinkevich_similarity_2007} proposed the following functions to parameterize the dimensionless gradients of velocity and potential temperature
\begin{align}
    \phi_m &= 1 + C_u \frac{z}{L_*}, \label{eq:most_extension_m}  \\ 
    \phi_h &= 1 + C_{H1}\!\left(\frac{z}{L_*}\right) + C_{H2}\!\left(\frac{z}{L_*}\right)^{2}. \label{eq:most_extension_h} 
\end{align}

In \cite{zilitinkevich_similarity_2007}, the empirical constants in equations \ref{eq:most_extension_c}, \ref{eq:most_extension_m}, and \ref{eq:most_extension_h} were specified as follows: $C_N = 0.1$, $C_f = 1$, $C_u = 2$, $C_{H1} = 1.6$, and $C_{H2} = 0.2$. Here, we recalibrate these constants based on our DNS data in accordance with the surface-layer scaling assumption. The resulting values are provided in the figure captions.

\begin{figure}
    \centering
    \includegraphics[width=\linewidth]{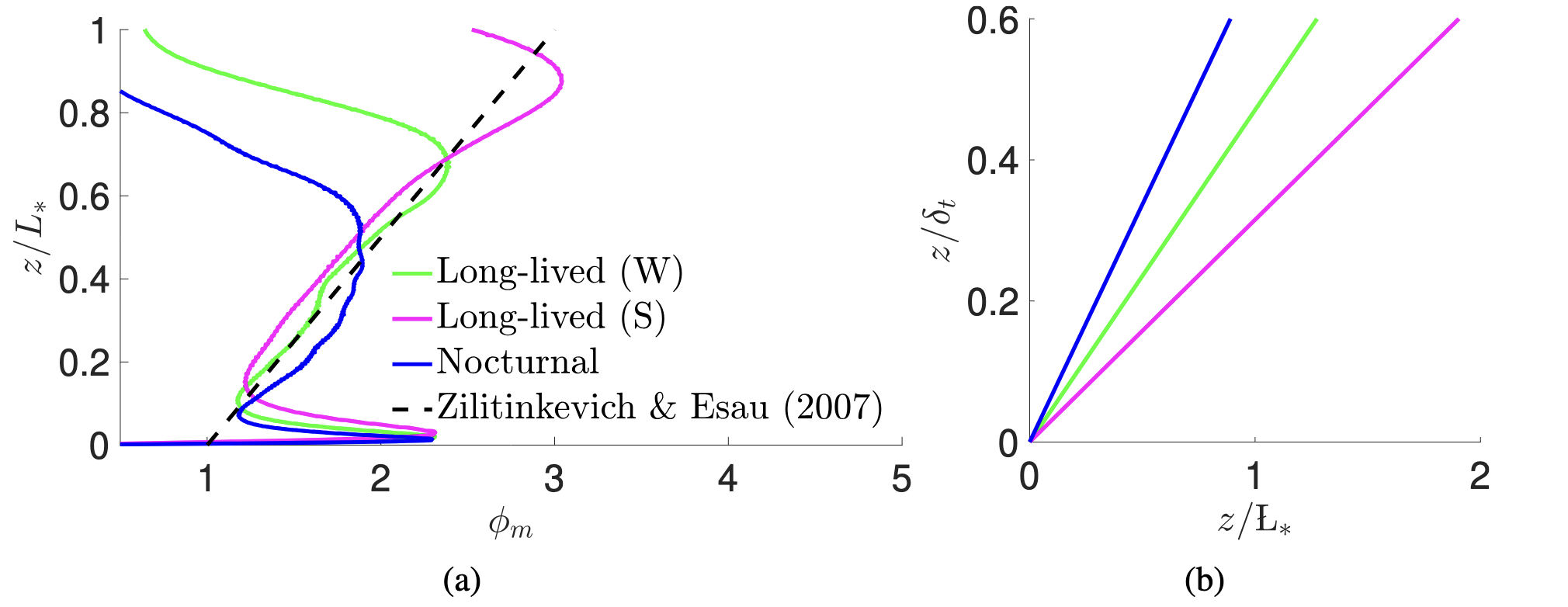}
    \caption{(a) Dimensionless velocity gradient $\phi_m$ versus $z/L_*$ for all three stratified cases. The dashed line represents equation \ref{eq:most_extension_m} with $C_u=2.0$, evaluated under the surface-layer scaling assumption in $L*$ with $C_N=0.19$ and $C_f=1.35$. (b) Relationship between $z/\delta_t$ and $z/L_*$.}
    \label{fig:most}
\end{figure}


Figure~\ref{fig:most} compares the DNS profiles of $\phi_m$ for all three stratified cases with the extended MOST function given by Equation \ref{eq:most_extension_m}, evaluated under the surface-layer scaling assumption used in the composite length scale $L_*$. Figure~\ref{fig:most}b shows the relationship between the dimensionless coordinates $z/L_*$ and $z/\delta_t$. 

Within the range of $z/\delta_t$ over which the Reynolds stresses and buoyancy fluxes decrease with distance from the surface (as shown in Figure~\ref{fig:most}b), all three stratified cases exhibit close agreement with the extended MOST function. This indicates that the formulation of \citeauthor{zilitinkevich_similarity_2007}, with its original constant $C_u=2.0$ and the recalibrated $L_*$ based on the surface-layer scaling assumption, is applicable to both nocturnal and long-lived stable atmospheric boundary layers (SABLs). The composite length scale $L_*$, through its dependence on $L_N = u_*/N_a$, effectively incorporates the influence of ambient stratification on the momentum field, such that a single slope captures profiles spanning weak and strong ambient stratification, as well as the nocturnal limit ($N_a = 0$). 

 
The range of applicability of the extended theory is bounded both from below and above, with both bounds showing a dependence on the ambient stratification. 
In the vicinity of the surface, viscous effects dominate and $\phi_m$ departs from the linear profile. This departure extends up to $z/L* \approx 0.08$ for the nocturnal case, $z/L* \approx 0.1$ for the long-lived (W) case, and $z/L* \approx 0.18$ for the long-lived (S) case. Stronger ambient stratification therefore raises the lower bound of applicability in dimensionless $z/L_*$ units. At the upper end, all three profiles curl away from the dashed line above $z/L_* \approx 0.6$--$0.7$, marking the end of the surface layer where similarity theory no longer holds. The upper limit itself shifts to higher $z/L_*$ as $N_a$ increases.

\begin{figure}
    \centering
    \includegraphics[width=0.60\linewidth]{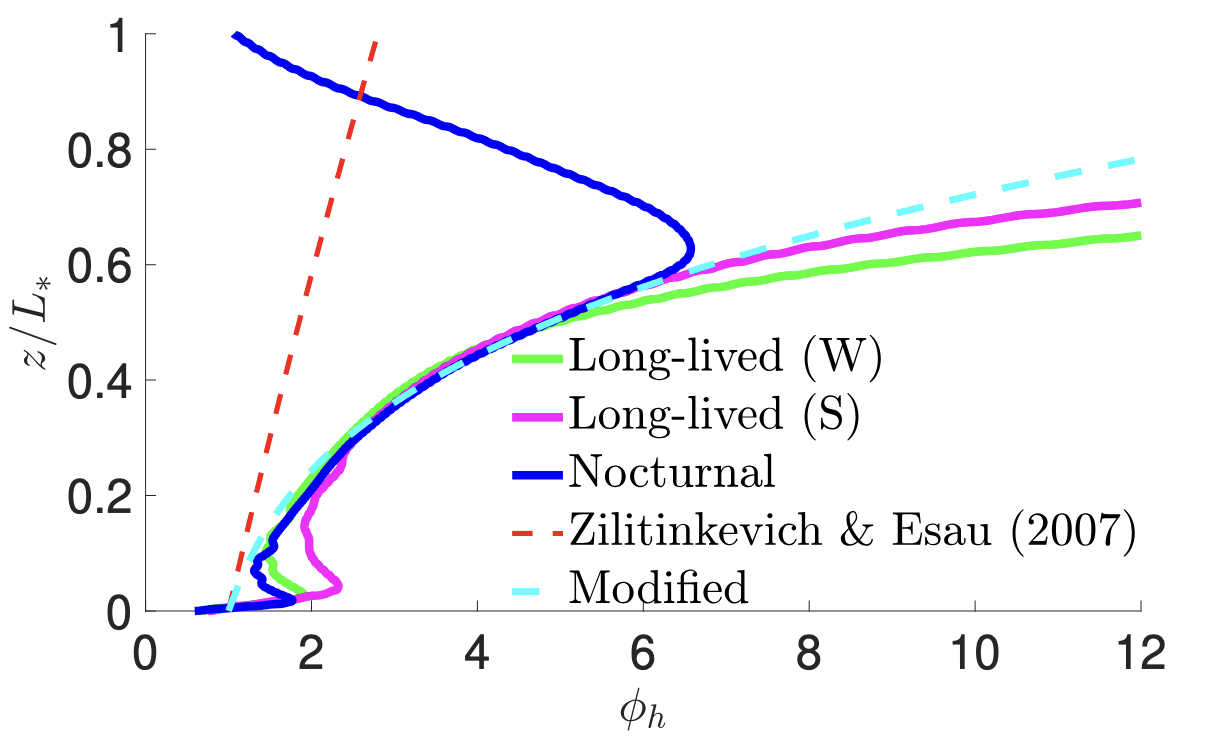}
    \caption{Dimensionless gradient of potential temperature $\phi_h$ versus $z/L_*$ for all three stratified cases. The dashed red line represents equation \ref{eq:most_extension_h} with $C_{H1}=1.6$ and $C_{H2}=0.2$, evaluated under the surface-layer scaling assumption in $L*$ with $C_N=0.19$ and $C_f=1.35$. The dashed cyan ``modified'' curve is equation~\ref{eq:modified_phih}.}
\label{fig:most_phi_h}
\end{figure}

Figure~\ref{fig:most_phi_h} presents the profiles of dimensionless gradients of potential temperature $\phi_h$. The composite length scale $L_*$, with its recalibrated constants ($C_N =0.19, C_f=1.35$) demonstrates a remarkable ability to collapse the $\phi_h$ profiles over the range $0.2\lesssim z/L_* \lesssim 0.6$ for all three stratified cases. The collapse of the profiles highlights the significant influence of ambient stratification --- absent in the nocturnal case --- on the composite length scale $L*$.  

Unlike the $\phi_m$ profiles, the extended MOST function given by equation \ref{eq:most_extension_h}, with its original empirical constants ($C_{H1} = 1.6$, $C_{H2} = 0.2$) and adopting the same $L_*$ with the recalibrated constants, underpredicts the DNS profiles for all three cases and fails to capture the shape of the profiles. To address this deficiency, we propose the following function to better represent the $\phi_h$ profiles:
\begin{equation}\label{eq:modified_phih}
    \phi_h = 1 + C_{H1}\!\left(\frac{z}{L_*}\right)
           + C_{H2}\!\left(\frac{z}{L_*}\right)^{2}
           + C_{H3}\!\left(\frac{z}{L_*}\right)^{3},
\end{equation}
with the constants calibrated to $C_{H1} = 2.5$, $C_{H2} = 3$, and $C_{H3} = 15$. 
This ``modified'' MOST function captures the DNS profiles of all three cases where there is a clear collapse of the curves, and it accommodates the strong upward sweep of $\phi_h$.

\begin{figure}
    \centering  
    \includegraphics[width=\linewidth]{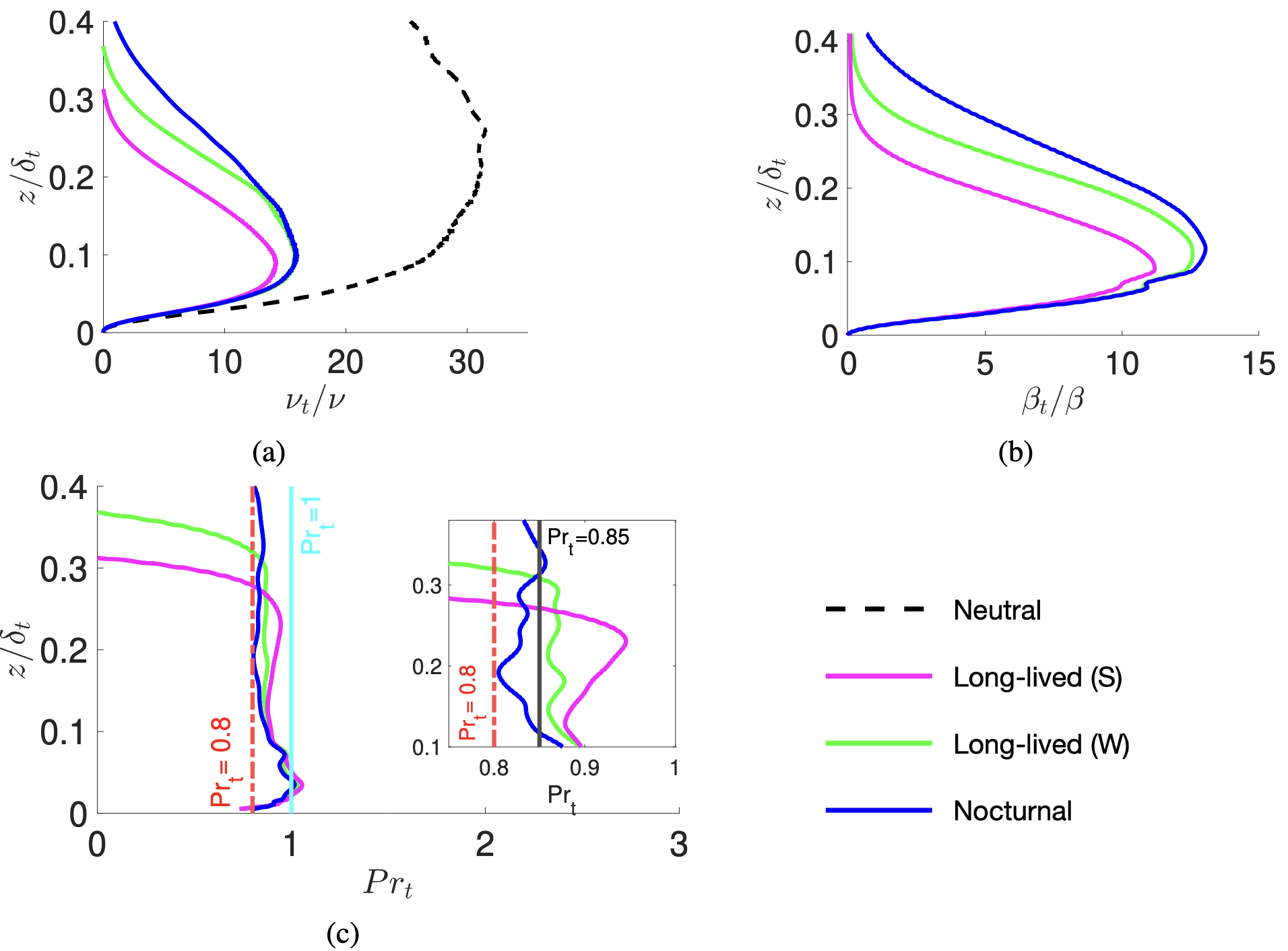}
    \caption{Effect of ambient stratification on the vertical profiles of (a)  turbulent Prandtl number ($Pr_t$), (b) eddy viscosity ($\nu_t$), and (c) eddy diffusivity ($\beta_t$).}
\label{fig:Prandtl_number}
\end{figure}

\subsection{Turbulent Prandtl number profiles}\label{sec:prandtlnumber}

A common approach in turbulence closure for scalar mixing in SABL is to prescribe the scalar eddy diffusivity as a constant multiple of the eddy viscosity, thereby assuming a fixed turbulent Prandtl number ($Pr_t$). In their DNS of nocturnal SABL up to $Re_D = 900$, \citet{shah_direct_2014} found that $Pr_t$ exhibits only a weak dependence on stability. Specifically, their results show profiles that peak near unity close to the surface and decrease approximately linearly to about 0.85 with increasing distance from the surface.

In contrast, our results for long-lived SABLs suggest that ambient stratification exerts a notable influence on TPE transport throughout the boundary layer, while its impact on TKE transport remains comparatively modest. These findings motivate a closer examination of the resulting $Pr_t$ profiles in long-lived SABLs.

We compute the eddy viscosity $\nu_t$, and diffusivity $\beta_t$ as follows
\begin{equation}
  \nu_t = - \frac{\langle u^\prime w^\prime \rangle \partial_z \langle u \rangle + \langle v^\prime w^\prime \rangle \partial_z \langle v \rangle}{(\partial_z \langle u \rangle)^2+(\partial_z \langle v \rangle)^2}, \qquad
\beta_t = -\frac{g\langle w^\prime \Phi^\prime \rangle}{g \partial_z \langle \Phi \rangle + N_a^2}.
\end{equation}
The turbulent Prandtl number is then calculated as
\begin{equation}
    Pr_t = \frac{\nu_t}{\beta_t}.
\end{equation}
We note that $N_a^2$ appears explicitly in $\beta_t$ to account for gradient of the background potential temperature profile. 

\begin{figure}
    \centering  
    \includegraphics[width=\linewidth]{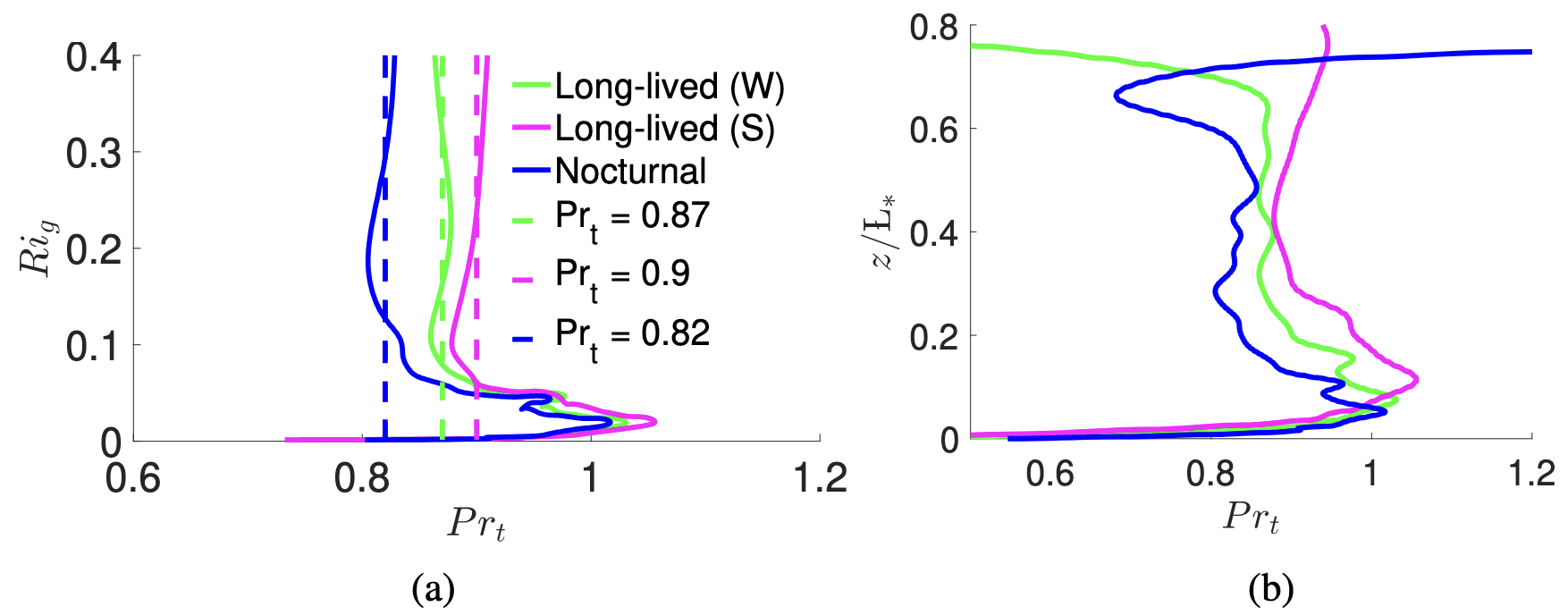}
    \caption{Dependence of $Pr_t$ on (a) $Ri_g$ and (b) $z/L_*$ across all three stratified cases.}
    \label{fig:prandtl_2}
\end{figure}

 Figure \ref{fig:Prandtl_number}(a) presents the eddy viscosity profiles for all three stratified cases along with the neutral case. The magnitudes of the eddy viscosity in all three stratified cases are significantly lower than in the neutral case, and decrease systematically with increasing ambient stratification. A similar trend is observed in the eddy diffusivity profiles shown in Figure \ref{fig:Prandtl_number}(b), where the magnitudes also diminish with increasing ambient stratification. The shape of these profiles and their response to increasing stability are consistent with the results reported by \citet{shah_direct_2014}.

 Figure \ref{fig:Prandtl_number}(c) presents the $Pr_t$ profiles for all three stratified cases. Focusing on the region $0.1 \lesssim z/\delta_t \lesssim 0.4$, where agreement with similarity theory is observed previously in figures \ref{fig:most} and \ref{fig:most_phi_h}, $Pr_t$ exhibits an approximately constant profile, bounded between 0.80 and 0.90. Within this region, its magnitude increases systematically with increasing ambient stratification. 

The Reynolds analogy suggests that turbulent transport of momentum and heat is governed by the same underlying mechanism, leading to a turbulent Prandtl number of $Pr_t = 1$. However, quantifying the vertical structure of $Pr_t$ in SABL has proven difficult, and no widely accepted model currently exists \citep{li_turbulent_2019}. To better understand the behavior of $Pr_t$ in long-lived SABLs, we examine its dependence on the gradient Richardson, $Ri_g$, and the composite stability parameter, $z/L_*$.  Accounting for the presence of ambient stratification, $Ri_g$ is computed as
\begin{equation}\label{eq:Ri_g}
    Ri_g = \frac{\partial \langle b \rangle/\partial z + N_a^2}{(\partial \langle u \rangle/\partial z)^2 + (\partial \langle v \rangle/\partial z)^2}.
\end{equation}

Figure~\ref{fig:prandtl_2}a presents $Ri_g$ as a function of $Pr_t$ for all three stratified cases. We exclude values of $Ri_g$ and $Pr_t$ when the corresponding stresses or fluxes are effectively zero, as these lead to spurious results. For $Ri_g \gtrsim 0.1$, all three cases show that $Pr_t$ approaches an approximately constant value that is largely independent of $Ri_g$. However, this constant differs among the cases and increases with the strength of ambient stratification: $Pr_t \approx 0.82$ for the nocturnal case, $Pr_t \approx 0.87$ for the long-lived (W) case, and $Pr_t \approx 0.90$ for the long-lived (S) case. Thus, while $Pr_t$ does not vary systematically with $Ri_g$, the value it approaches depends on the ambient stratification rather than being universal.

Figure \ref{fig:prandtl_2}b presents the variation of $Pr_t$ with the composite stability parameter $z/L_*$. The influence of ambient stratification is consistent with the trends observed in the previous figures. Over the range $z/L_* \approx 0.1$ to $z/L_* \approx 0.5$, $Pr_t$ decreases slightly with increasing stability parameter. Due to the relatively low $Re_D$ of the present DNS data, it remains difficult to draw definitive conclusions regarding the vertical structure of $Pr_t$. Nonetheless, these results suggest that future studies should explicitly account for the role of ambient stratification.




\subsection{Regime classification of long-lived SABL}\label{sec:regime}

\begin{table}
\footnotesize
\caption{Simulation parameters used for regime classification of long-lived SABL at fixed Reynolds number. From left: Case number, dominant stratification mechanism (DSM), $\Pi_w$, $\Pi_s$, Reynolds number, domain size, number of grid points in $x-y$, and $z$ directions, bulk $(Ri_b)$ and surface $(Ri_s)$ Richardson numbers. The first three cases (I-III) are used to show different dynamics for the same $Ri_s$ in Section \ref{sec:regime_1}.}  
\label{tab:stratifiedparam} 
\centering
\begin{tabular}{@{}lllcccccc@{}}
\toprule
    Cases & DSM & $\Pi_w$ & $\Pi_s$ & $Re_{D}$& Domain Size& $N_x^2 \times N_z$ & $Ri_b$ & $Ri_s$  \\  \midrule   
I \ & $N_a$ dom.           & 4050      & 0.25    & 400  &$26D \times 26D \times 24D$&   $128^2 \times 512$ & 2.0607 &  0.4121  \\
II \ & $B_s \ \& \ N_a$    & 8100      & 1       & 400  &$26D \times 26D \times 24D$&   $128^2 \times 512$ & 0.8243 &  0.4121  \\
III \ & $B_s$ dom.         & 16200     & 4       & 400  &$26D \times 26D \times 24D$&   $128^2 \times 512$ & 0.5152 &  0.4121  \\ \hline  
IV \ & $N_a$ dom.          & 16200     & 0.25    & 400  &$26D \times 26D \times 24D$&   $128^2 \times 512$ & 0.1288 &  0.0258  \\
V \ & $B_s \ \& \ N_a$     & 16200     & 1       & 400  &$26D \times 26D \times 24D$&   $128^2 \times 512$ & 0.2061 &  0.1030  \\
VI \ & $B_s$ dom.          & 16200     & 8       & 400  &$26D \times 26D \times 24D$&   $128^2 \times 512$ & 0.9273 &  0.8243  \\
VII \ & $B_s$ dom.         & 16200     & 10      & 400  &$26D \times 26D \times 24D$&   $128^2 \times 512$ & 1.1334 &  1.0303  \\
VIII \ & $B_s$ dom.        & 16200     & 16      & 400  &$26D \times 26D \times 24D$&   $128^2 \times 512$ & 1.7516 &  1.6485  \\
IX \ & $B_s$ dom.          & 12810     & 4       & 400  &$26D \times 26D \times 24D$&   $128^2 \times 512$ & 0.8239 &  0.6591  \\  
X \ & $N_a$ dom.           & 8100      & 0.25    & 400  &$26D \times 26D \times 24D$&   $128^2 \times 512$ & 0.5152 &  0.1030  \\
XI \ & $N_a$ dom.          & 8100      & 0.5     & 400  &$26D \times 26D \times 24D$&   $128^2 \times 512$ & 0.6182 &  0.2061  \\
XII \ & $B_s$ dom.         & 8100      & 4       & 400  &$26D \times 26D \times 24D$&   $128^2 \times 512$ & 2.0607 &  1.6485  \\
XIII \ & $N_a$ dom.        & 6404      & 0.25    & 400  &$26D \times 26D \times 24D$&   $128^2 \times 512$ & 0.8242 &  0.1648  \\
XIV \ & $N_a$ dom.         & 4050      & 0.0625  & 400  &$26D \times 26D \times 24D$&   $128^2 \times 512$ & 1.7516 &  0.1030  \\
XV \ & $N_a$ dom.          & 4050      & 0.125   & 400  &$26D \times 26D \times 24D$&   $128^2 \times 512$ & 1.8546 &  0.2061  \\
XVI \ & $B_s \ \& \ N_a$   & 4050      & 1       & 400  &$26D \times 26D \times 24D$&   $128^2 \times 512$ & 3.2971 &  1.6485  \\
XVII \ & $B_s$ dom.        & 4050      & 4       & 400  &$26D \times 26D \times 24D$&   $128^2 \times 512$ & 8.2426 &  6.5941  \\ \hline
  \rowcolor{gray!20}XVIII \ & $N_a$ dom.  & 6404  & 0.25        & 400  &$104D \times 104D \times 24D$&   $512^2 \times 512$ & 0.8242 &  0.1648\\
 \rowcolor{gray!20}XIX \ & $B_s$ dom.            & 16200        & 4        & 400  &$104D \times 104D \times 24D$&   $512^2 \times 512$ & 0.5152 &  0.4121  \\
 \hline 

\end{tabular}
\end{table}
Thus far, we have seen that long‑lived SABLs differ markedly from their nocturnal counterparts, particularly in the structure of their thermal fields. In \S\ref{sec:numerics_nondim_numbers}, we established an expanded dimensionless parameter space for long‑lived SABLs. Because ambient stratification is explicitly included, this parameter space contains one additional degree of freedom relative to that of a nocturnal SABL.

Nocturnal SABLs are commonly categorized into weakly stable and very stable regimes following the classification originally proposed by \citet{mahrt_stratified_1998}. 
In this framework, the weakly stable regime is characterized by turbulence that is continuous in space and time, whereas the very stable regime exhibits intermittent turbulence and global intermittency \citep{mahrt_stably_2014}. 
In this section, we assess whether this classification framework extends to long-lived SABLs despite the presence of ambient stratification and examine how these two regimes manifest within the expanded parameter space characteristic of long-lived conditions. To this end, we construct a regime map that delineates the conditions under which distinct turbulent behaviors arise in long-lived SABLs, thereby elucidating the dependence of flow regimes on the governing dimensionless parameters. 

Following \citet{ren_research_2025}, we adopt the onset of turbulence collapse induced by stable stratification as a qualitative indicator of the very stable regime. Systematically mapping this transition across the relevant dimensionless parameter space requires a large ensemble of DNS. However, the substantial computational cost associated with each simulation renders exhaustive exploration at high $Re$ impractical. To balance physical fidelity with the cost of performing DNS, we therefore fix the Reynolds number at $Re_D = 400$, which allows us to efficiently survey the parameter space while retaining the essential dynamics of stratification-induced turbulence suppression. In total, nineteen DNS are conducted, with the parameters of each simulation summarized in Table~\ref{tab:stratifiedparam}.

The use of a relatively low $Re$ affords an additional methodological advantage. At such $Re$, the flow remains amenable to linear stability analysis, which provides an efficient and exact means of identifying regions of parameter space that are linearly stable and incapable of sustaining turbulence. This enables a clear delineation of the theoretical boundary beyond which turbulence collapse must occur. The combined use of DNS and linear stability analysis therefore forms a complementary framework that would not be simultaneously tractable at higher $Re$. Their joint deployment allows for a more comprehensive characterization of the regime map than either approach could provide in isolation. Full details of the linear stability analysis and its validation are presented in Appendix~\ref{sec:appendix_lsa}.

With the $Re$ fixed, the ranges of the remaining governing dimensionless parameters are selected to span and demarcate the linearly stable, weakly stable, and very stable regimes. Complete flow laminarisation is observed for $\Pi_w \leq 4050$, illustrating the sensitivity of the regime boundaries to the stratification parameters. Near-wall resolution is maintained at $\Delta_x^+ (= \Delta_y^+) = 5.97$ and $\Delta_z^+ = 1.21$, consistent with the $Re_D$ case reported by \citet{shah_direct_2014}, ensuring that the near-wall turbulence dynamics are adequately resolved. The influence of domain size on regime manifestation is assessed in appendix~\ref{sec:appendix_domainsize}, where it is demonstrated that the present domain is sufficient for the time-series analysis employed for regime identification.

In what follows, we first demonstrate the need for a multi-parameter space for regime classification in long-lived SABL, followed by the effect of $\Pi$-parameters on turbulence collapse, which serves as a qualitative indicator for characterizing long-lived SABL. The section concludes with a regime map delineating long-lived SABL into linearly stable, weakly stable, and very stable regimes.

\subsubsection{Dependence on dimensionless parameter space }\label{sec:regime_1}

\begin{figure}
\centering
    \includegraphics[width=\linewidth]{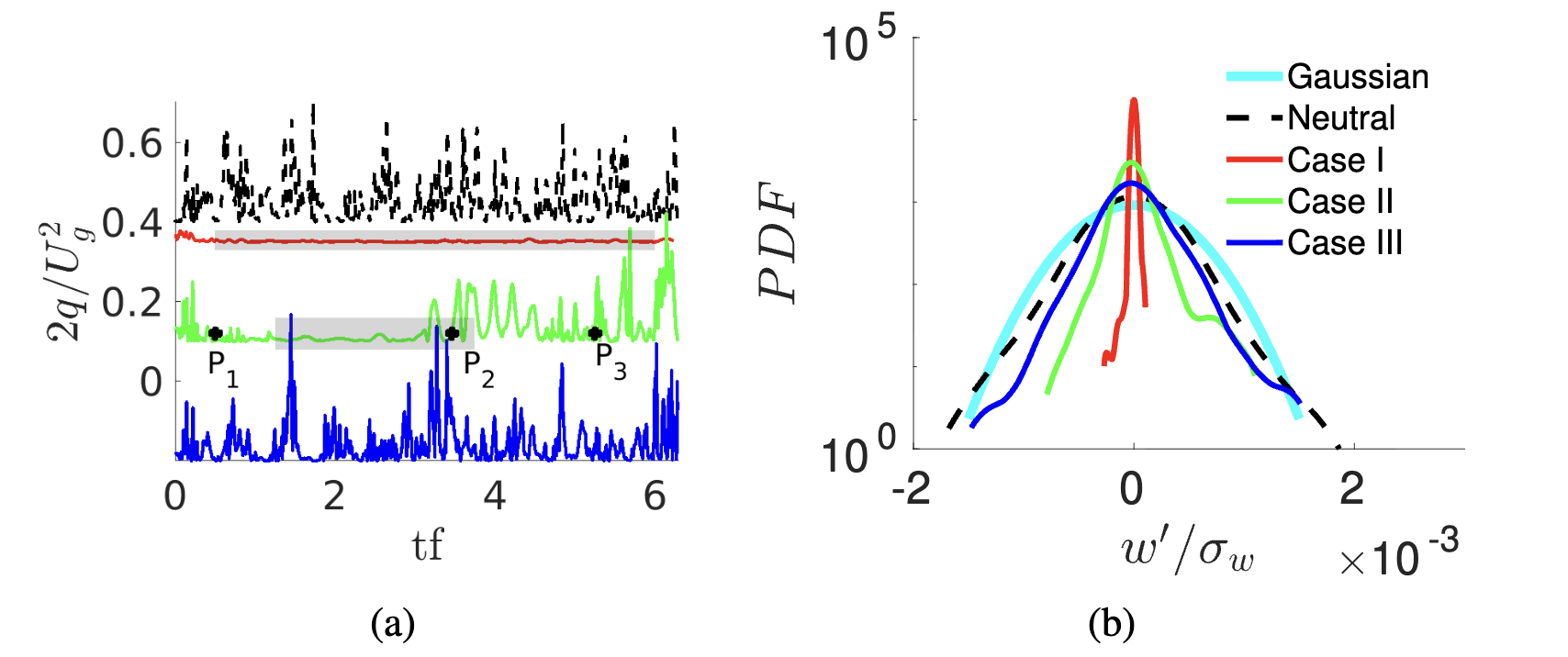}   
    \caption{Effect of $\Pi_s$ and $\Pi_w$ on (a) the TKE time series and (b) the PDF of vertical-velocity fluctuations, both sampled at the geometric centre of a horizontal plane at $z/D\,(z^{+}) = 0.5\,(32.70)$. In panel (a), the grey shaded region denotes the period of turbulence collapse; markers $P_1$–$P_3$ indicate time instances immediately before collapse, during collapse, and after turbulence resurgence, respectively; and the black, red, green, and blue curves are shifted vertically by $0.4$, $0.35$, $0.1$, and $-0.2$, respectively, for visual clarity. In panel (b), the cyan curve shows the analytical Gaussian distribution with zero mean and unit standard deviation.}
    \label{fig:Ri_timeseries}
\end{figure}

\begin{figure}
\centering
        \includegraphics[width=\textwidth]{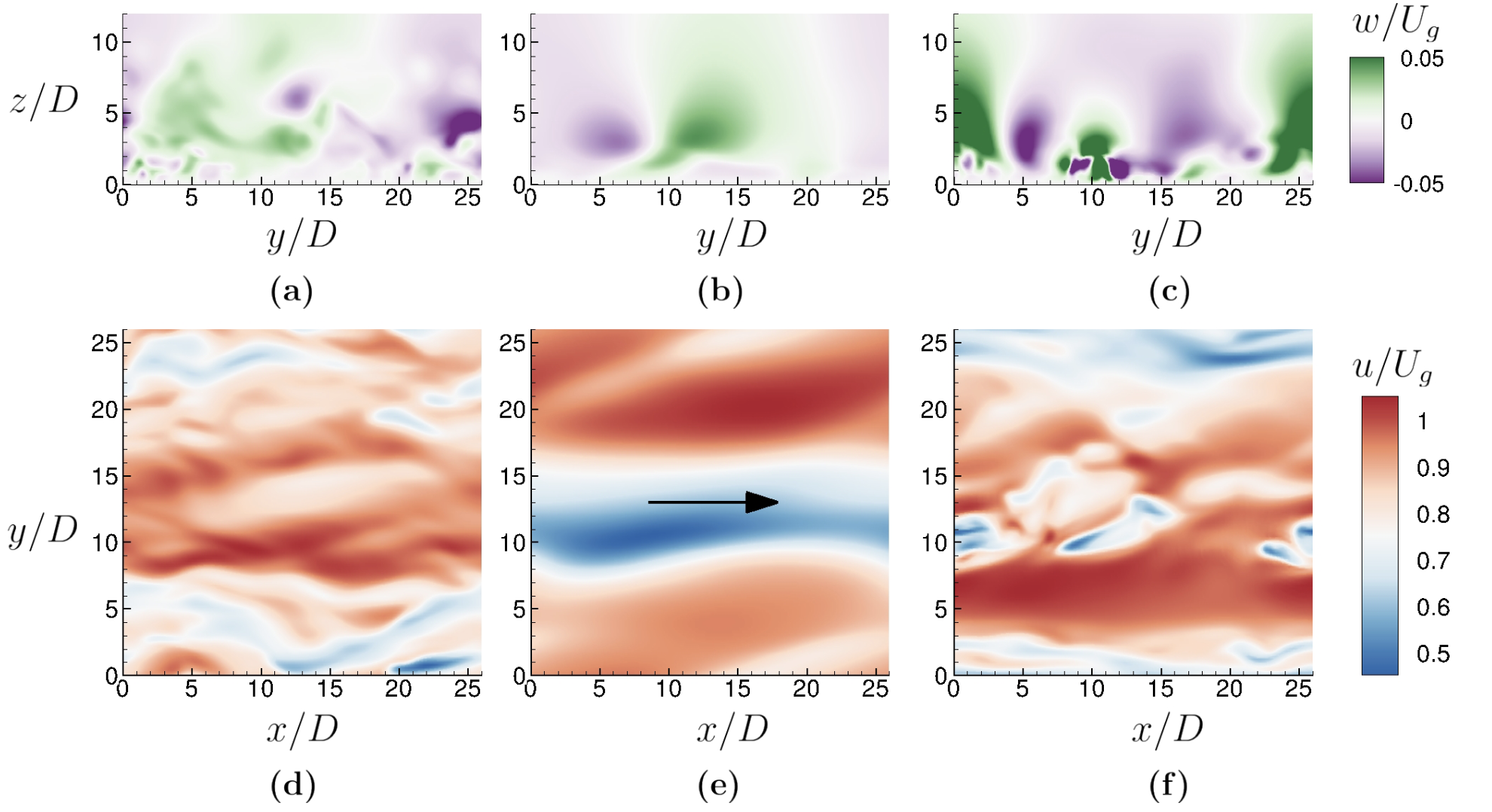}
    \caption{Instantaneous contours of vertical (a–c) and horizontal (d–f) velocities for Case~\Romannum{2} ($\Pi_w = 8100$, $\Pi_s = 1$) in the $y$–$z$ and $x$–$y$ planes, respectively, at time instances $P_1$ (a,d), $P_2$ (b,e), and $P_3$ (c,f) marked in figure~\ref{fig:Ri_timeseries}(a). The arrow indicates the geostrophic wind direction.}
    \label{fig:Ri_instantfield}
\end{figure}
For fixed Reynolds and Prandtl numbers, an idealized nocturnal SABL can be characterized by a single dimensionless parameter, typically the Richardson number \citep{coleman_direct_1992, shah_direct_2014, deusebio_numerical_2014, ansorge_global_2014}. By adopting an idealized configuration, we deliberately omit the additional complexities of real SABLs that would otherwise enlarge the dimensionless parameter space. In contrast, an idealized long-lived SABL involves two independent stratification mechanisms; consequently, even at fixed $\Rey$ and $\Pran$, the flow depends on two additional dimensionless parameters. This expanded parameter space fundamentally distinguishes long-lived SABLs from their nocturnal counterparts.

To examine how flow regimes in long-lived SABLs depend on this expanded set of dimensionless parameters, we consider cases \Romannum{1}–\Romannum{3} in Table~\ref{tab:stratifiedparam}. These cases share the same surface Richardson number ($Ri_s = 0.04$) but differ in their values of $\Pi_s$ and $\Pi_w$, thereby isolating the effects of the additional stratification parameters.

Probability density functions (PDF) of intermittent turbulence are characterised by a sharp peak centred at zero, reflecting extended quiescent periods, and heavy tails corresponding to rare intense events \citep{Lohse1993}. In grid turbulence, PDFs of velocity fluctuations transition from sub-Gaussian in developing turbulence, to Gaussian in fully developed turbulence, to hyper-Gaussian in decaying turbulence \citep{mouri_2002}. We exploit deviation from Gaussian behaviour as a qualitative diagnostic for the very stable regime, where turbulence undergoes intermittent growth and collapse.

Figure~\ref{fig:Ri_timeseries}(a) shows the time evolution of the vertical velocity at the centre of the $x$–$y$ plane at $z^{+} = 32$. In Case~\Romannum{1} ($\Pi_w = 4050$), turbulence originating from the neutrally stratified initial condition collapses shortly after stratification is introduced, whereas in Case~\Romannum{2} ($\Pi_w = 8100$) it persists for a longer period before decaying over an extended interval and subsequently re-emerging, as indicated by the $P_3$ marker. In Case~\Romannum{3} ($\Pi_w = 16200$), by contrast, turbulence remains sustained throughout the simulation.

Figure~\ref{fig:Ri_timeseries}(b) shows PDF of vertical-velocity fluctuations for the three cases of panel (a). Progressing from Case~\Romannum{1} to \Romannum{3}, the PDFs broaden systematically from a narrow, highly peaked distribution with short tails to a wider, Gaussian-like form with wider tails, reflecting the transition from complete turbulence collapse to sustained turbulence. This interpretation is corroborated by the close agreement between the Case~\Romannum{3} PDF and the analytical Gaussian, consistent with the continuous, well-developed turbulence observed in panel (a). 

Figure~\ref{fig:Ri_instantfield} shows instantaneous contours of the vertical (a–c) and horizontal (d–f) velocity fields for Case~\Romannum{2} ($\Pi_w = 8100$, $\Pi_s = 1.0$) at time instances $P_1$–$P_3$ marked in figure~\ref{fig:Ri_timeseries}(a). The vertical velocity fields capture the transition from a turbulent state at $tf = 0.5$ ($P_1$) to a dynamically unstable, non-turbulent state at $tf = 3.46$ ($P_2$), and the subsequent resurgence of turbulence at $tf = 5.25$ ($P_3$). The horizontal velocity fields (d–f) exhibit a corresponding evolution: during turbulence collapse, alternating low- and high-momentum bands emerge, which are then disrupted upon turbulence resurgence and accompanied by the formation of secondary structures, as evident in panel (f).



\subsubsection{Impact on sustenance of turbulence}\label{sec:regime_2}

\begin{figure}
\centering
    \includegraphics[width=0.9\linewidth]{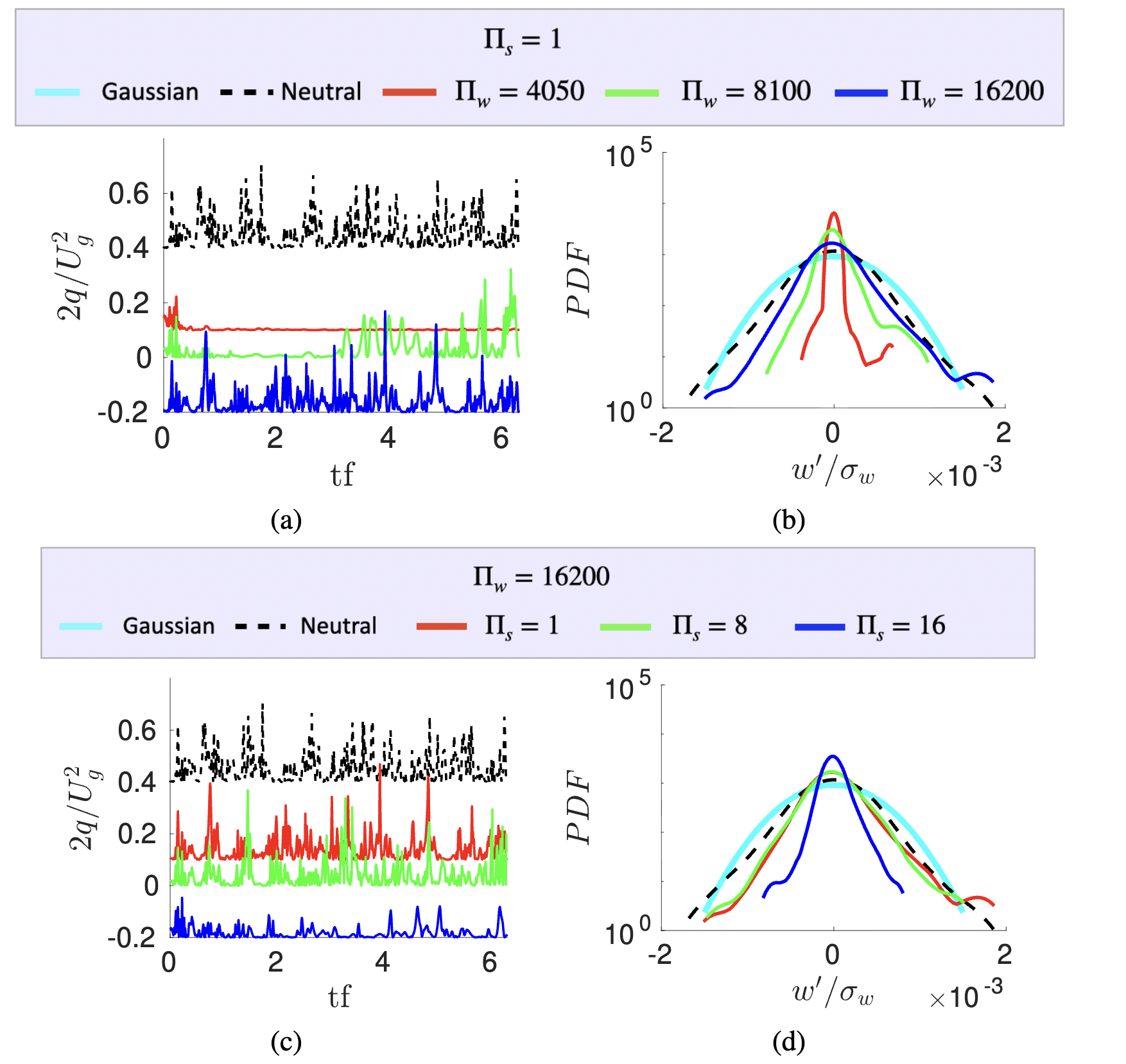}   
    \caption{Effect of $\Pi_w$ ($= 4050, 8100, 16200$) at fixed $\Pi_s = 1.0$ (a,b), and of $\Pi_s$ ($= 1, 8, 16$) at fixed $\Pi_w = 16200$ (c,d), on (a,c) the time evolution of TKE ($q$) and (b,d) the PDF of vertical-velocity fluctuations. All data are sampled at the centre of the horizontal plane at $z/D = 0.5$ ($z^+ = 32$). In panels (a,c), TKE time series are shifted upwards by $0.1$, $0.2$, and $0.3$ relative to the neutrally stratified case (black dashed line), and the marker denotes the time instance of turbulence collapse. In panels (b,d), the cyan curve shows the analytical Gaussian distribution with zero mean and unit standard deviation.}
    \label{fig:timeseries}
\end{figure} 

We start with examining the influence of the $\Pi_w$ parameter on turbulence sustenance. 
Figure~\ref{fig:timeseries}(a–b) illustrates the effect of varying $\Pi_w$ ($4050$, $8100$, and $16200$) while keeping $\Pi_s = 1.0$ fixed. As shown in panel (a), a clear transition in flow regime occurs with decreasing $\Pi_w$. For $\Pi_w = 16200$, turbulence is continuously sustained, indicating a weakly stable regime. At $\Pi_w = 8100$, intermittent collapse and resurgence of turbulence signal a transition toward a very stable regime. For the lowest value, $\Pi_w = 4050$, turbulence ceases entirely, confirming a very stable regime.  
The PDFs of $v'$ in panel (b) exhibit a similar trend, with the distribution widening toward a Gaussian reference as $\Pi_w$ increases. Similar behavior is observed for $\Pi_s = 0.25$ (ambient-stratification dominated) and $\Pi_s = 4.0$ (surface-stratification dominated), where turbulence collapse occurs at lower $\Pi_w$ values, while continuous turbulence is maintained for the strongest $\Pi_w$. These results demonstrate that increasing $\Pi_w$ shifts the flow toward a weakly stable regime with sustained turbulence.

Next, we examine the effect of $\Pi_s$ ($1.0$, $8.0$, and $16.0$) for a fixed $\Pi_w = 16200$, as shown in figure~\ref{fig:timeseries}. In panel (a), continuous turbulence is observed throughout the inertial cycle for $\Pi_s = 1.0$, whereas turbulence collapse and subsequent resurgence occur frequently for $\Pi_s = 8.0$. Further increase to $\Pi_s = 16.0$ results in a prolonged turbulence collapse before fluctuations eventually resume.  These observations indicate that increasing $\Pi_s$ drives the flow from a weakly stable to a very stable regime. The PDFs of $v'$ in panel (b) exhibit a similar trend, with the distribution widening as $\Pi_s$ decreases from $16.0$ to $1.0$. For sufficiently large $\Pi_w$, where the flow is inertia dominated, a very stable regime can still emerge due to strong surface cooling, as reflected by a high $\Pi_s$. This behavior persists for $\Pi_w \gtrsim 8100$, whereas for lower $\Pi_w$, the ambient stratification is strong enough to suppress turbulence resurgence throughout the inertial cycle, independent of $\Pi_s$ (not shown here).

\begin{figure}
    \centering
    \includegraphics[width=0.95\linewidth]{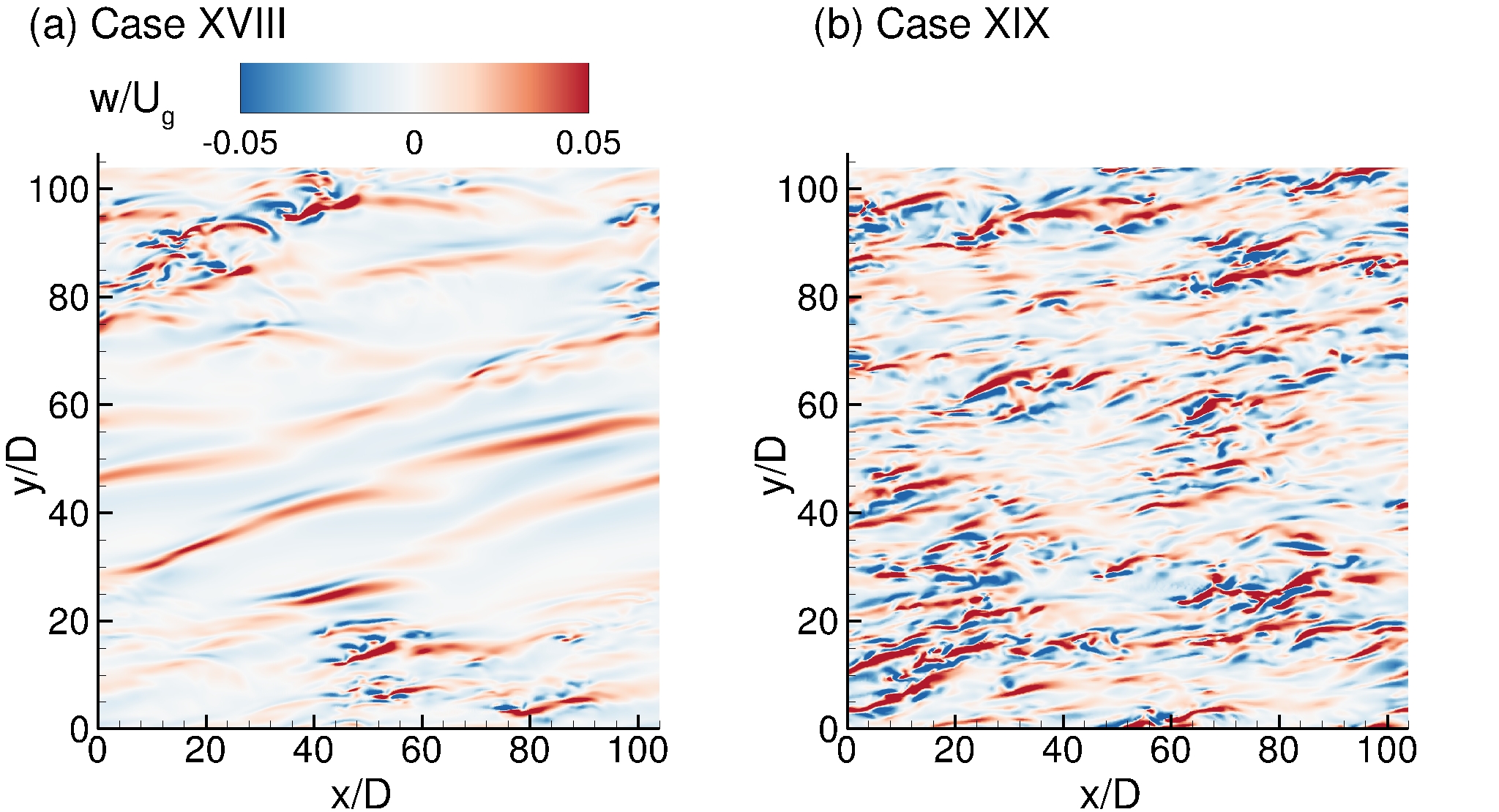}
    \caption{Instantaneous vertical velocity at $z/D (z^+) = 0.25 (17))$ for (a) $N_a$ dominant (Case~\Romannum{18}) and (b) $B_s$ dominant (Case~\Romannum{19}) cases.}
    \label{fig:wsabl_vsabl}
\end{figure}

Figure \ref{fig:wsabl_vsabl} shows contour visualization of instantenous vertical velocity corresponding to the ambient dominant (case XVIII) and surface cooling dominant (case XIX) cases from table \ref{tab:stratifiedparam}.  The two cases show the evolution of flow structures from a very stable regime towards a weakly stable one. Evidently, a strong turbulence collapse that causes relaminarization is the characteristics of the very stable regime (see panel a). As observed in figure \ref{fig:timeseries}, the effect of relaminarization can lasts for a very long time in long-lived SABL. On the contrary, as shown in panel (b), the flow exhibit small structures and enhanced turbulence as the flow approaches a weakly stable regime. It is important to note that at higher $\Rey_D$, we have observed intense turbulence in the weakly stable long-lived SABL, as shown in figure \ref{fig:contourplots}.

\begin{figure}
\centering
        \includegraphics[width=0.85\textwidth]{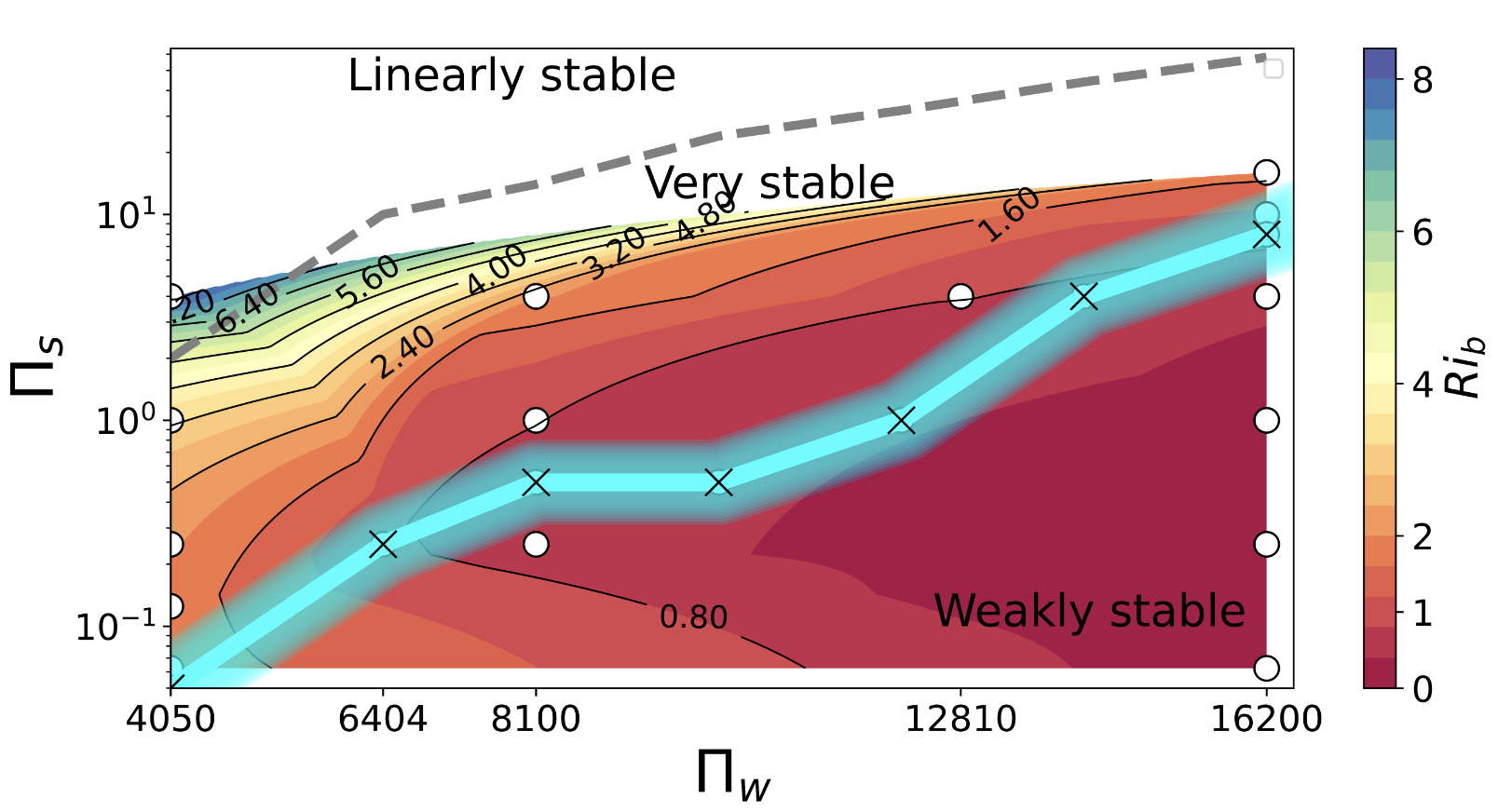}
        \caption{Regime map for the long-lived SABL at $Re_D=400$. Color represents contours of bulk Richardson number $(Ri_b)$ computed based on surface buoyancy flux and ambient stratification $N_a$ (see equation \ref{eq:conv_param}. Cyan color delineates the very stable and weakly stable regimes, whereas the dashed line represents the neutral curve obtained from linear stability analysis. 
        }
        \label{fig:regimes}
\end{figure}

\subsubsection{Regime map}\label{sec:regime_3}
Drawing on the DNS results of table~\ref{tab:stratifiedparam} and the linear stability analysis of appendix~\ref{sec:appendix_lsa}, we construct a regime map for the long-lived SABL at $Re_D = 400$, shown in figure~\ref{fig:regimes}. The map delineates the approximate boundaries between the weakly stable, very stable, and linearly stable regimes, drawn broadly to reflect the inherent uncertainty in these boundaries arising from the possible coexistence of subcritical turbulent and laminar states \citep{brethouwer2012turbulent}.

The regime map reveals a systematic progression from the linearly stable regime toward the weakly stable regime as $\Pi_s$ decreases and $\Pi_w$ increases. The weakly stable regime emerges at low $\Pi_s$, which may reflect either reduced surface cooling or enhanced ambient stratification. Although the promotion of sustained turbulence under stronger ambient stratification may appear counterintuitive, turbulence driven by ambient stratification behaves fundamentally differently from that driven by surface cooling alone \citep{xiao_impact_2022}; a low-$\Pi_s$ weakly stable regime therefore reflects the combined influence of both mechanisms. The map further shows that at larger $\Pi_w$, the weakly stable regime spans a wider range of $\Pi_s$, with this range narrowing as $\Pi_w$ decreases, highlighting the comparatively stronger role of ambient stratification relative to surface cooling in sustaining turbulence in a long-lived Ekman layer. Finally, the case with $Ri_b = 0.8$ straddles both the very stable and weakly stable regimes, underscoring the necessity of a multi-parameter framework for accurately characterising long-lived SABLs.

The transition from the very stable regime toward the linearly stable regime is complex. The linearly stable regime is broadly associated with large $\Pi_s$ and low $\Pi_w$, with its boundary defined by the neutral stability curve shown as a dashed line in figure~\ref{fig:regimes}. Within this regime, infinitesimal disturbances decay and the Ekman layer remains laminar in the linear approximation. Beyond the neutral curve, approaching the very stable regime, Type~\Romannum{2} instability emerges \citep{lilly_instability_1966,brown_inflection_1972}. A secondary stability analysis could further subdivide the very stable regime into a dynamically unstable but non-turbulent region and a regime characterized by intermittent turbulence; however, this was not pursued owing to the substantial computational cost involved.

Although the regime map is constructed for $Re_D = 400$ to maintain manageable computational cost, it provides a useful framework for interpreting the response of a long-lived SABL to variations in $\Pi_s$ and $\Pi_w$, and for identifying regions of parameter space where weakly stable turbulence is likely to occur. We expect this qualitative progression to hold at higher Reynolds numbers as well. This is corroborated at $Re_D = 900$, where a substantial reduction in both $\Pi_w$ (from $47527$ to $25775$) and $\Pi_s$ (from $16$ to $4$) follows the nonlinear demarcation curve yet preserves the weakly stable regime, confirming that the regime boundaries retain their qualitative structure at higher Reynolds numbers. The constant contours of the bulk Richardson number overlayed on the map show same contours spanning different regimes, highlighting the limitations of a single Richardson number as a quantitative indicator of stratification strength or metric for regime identification; nevertheless, from a qualitative perspective, an increase in the Richardson number at fixed values of the remaining dimensionless parameters generally corresponds to a progression toward a more stable regime, irrespective of its absolute value.

\section{Conclusions}\label{sec:conclusions}
We presented direct numerical simulations (DNS) of long-lived stable atmospheric boundary layers (SABLs) in which surface cooling and a constant uniform ambient stratification acted independently on the flow field. By contrasting cases of weak and strong ambient stratification at $Re_D=900$ against a nocturnal SABL without ambient stratification, we examined how the two stability mechanisms jointly governed turbulent transport and modified the similarity theory and scaling relationships crucial for parameterizing turbulent mixing in the SABL.

A striking feature of the simulated SABLs was the emergence of a multi-layered thermal structure driven by ambient stratification: a stable layer near the surface, an intermediate layer of reduced static stability, and an overlying stable layer capped by a buoyancy inversion. This inversion was absent in the initial profiles and grew increasingly pronounced with ambient stratification strength, indicating enhanced downward mixing of warmer air from aloft. Crucially, however, the reduced static stability of the intermediate layer was not caused by enhanced turbulent mixing — vertical profiles of turbulent kinetic energy and Reynolds stresses showed a monotonically decreasing trend with increasing ambient stratification strength, pointing instead to a stratification-driven reorganization of the thermal structure.

In accordance with this observation, we examined the turbulent kinetic energy (TKE) and turbulent potential energy (TPE) budgets for both long-lived and nocturnal SABLs. The budget analyses revealed that ambient stratification exerted a weak influence on the momentum field while strongly modifying the buoyancy field — a contrast corroborated by the TKE budgets, which remained largely insensitive to the stratification mechanism. The TPE, a measure of buoyancy variance, exhibited a markedly amplified peak within the intermediate layer, whose formation is explained by an additional production term proportional to the ambient stratification strength $N_a$. Near the surface, turbulence remained governed primarily by surface cooling, with ambient stratification exerting increasing influence within the intermediate layer, modulating both TPE production and dissipation. The buoyancy production associated with $N_a$ dominated the vertical structure, while production due to the mean buoyancy gradient reversed sign across the multi-layered thermal structure — transitioning from a source to a sink — to maintain energy balance. Collectively, these results reveal a TPE-centred transport mechanism unique to long-lived SABLs and lend strong support to turbulence closure approaches based on total-energy formulations \citep{zilitinkevich_turbulence_2008, Zilitinkevich2007_EFB, Zilitinkevich2012}.

The DNS of all three stratified cases were used to assess an extension of the Monin–Obukhov similarity functions proposed by \cite{zilitinkevich_similarity_2007}, in which a composite length scale $L_*$ incorporates contributions from ambient stratification and Earth's rotation alongside the classical Obukhov length scale. Recalibration of the empirical constants under the surface-layer scaling assumption demonstrated that this extended framework has strong potential to represent dimensionless gradients of both velocity, $\phi_m$, and potential temperature, $\phi_h$ --- the latter of which has historically proved particularly challenging. Notably, $\phi_h$ profiles collapsed excellently over the region where similarity theory is expected to hold when plotted against the composite stability parameter $z/L_*$. In accordance with this trend, we proposed a new similarity relationship that captures the variation of $\phi_h$ as a function $z/L_*$ with cubic order.

Vertical profiles of the turbulent Prandtl number exhibited a slight decreasing trend with increasing $z/L_*$, though not to a degree that challenges the constant $Pr_t$ assumption. Nevertheless, the approximately constant value of $Pr_t$ showed a clear dependence on ambient stratification strength, increasing from $Pr_t \approx 0.82$ in the nocturnal case to $Pr_t \approx 0.90$ in the long-lived SABL with the strongest ambient stratification.

The presence of dual stratification mechanisms naturally expands the governing dimensionless parameter space from three for a nocturnal SABL to four for a long-lived SABL (see equation \ref{eq:dim_param}). Building on this expanded parameter space, we combined DNS with linear stability analysis to construct a regime map in the $\Pi_s$--$\Pi_w$ plane at fixed Reynolds and Prandtl numbers. Following the classification of \citet{mahrt_stratified_1998}, the map delineated three regimes: a linearly stable regime in which the flow remained laminar, a weakly stable regime in which turbulence persisted continuously in space and time, and a very stable regime in which intermittent turbulence collapse and resurgence were observed. Increasing $\Pi_s$ drove the flow toward the linearly stable and very stable regimes, while increasing $\Pi_w$ promoted weakly stable turbulence. The map further suggested a practical guideline: although no universal threshold exists, lower bulk Richardson numbers at a fixed Reynolds number are consistently associated with a greater likelihood of sustained turbulence. Collectively, these findings provide compelling evidence for the inherently multi-parameter nature of the long-lived SABL and caution that subgrid-scale parameterisations relying on a single stratification parameter — such as the gradient Richardson number — may prove inadequate in practice, in line with prior guidance \citep{galperin_critical_2007, grachev_critical_2013}.

The present study underscored the significant influence of ambient stratification on the thermal structure of the SABL. Surface flux heterogeneity has also been shown to substantially alter the thermal structure of SABL, with increasing thermal contrast between patches exerting a particularly strong influence \citep{stoll2009}. A natural extension of the present work would be to introduce heterogeneous surface cooling in the form of alternating patches within the long-lived SABL framework.

\appendix
\section{Effect of horizontal domain extent on statistics}\label{sec:appendix_domainsize}
Inadequately sized computational domains can generate spurious flow structures in direct numerical simulations of stratified atmospheric boundary layer flows as discussed in prior works \citep{coleman_direct_1992, garcia-villalba_turbulence_2011, deusebio_numerical_2014}. To ensure that the present results are not influenced by domain-size limitations, a detailed sensitivity analysis was conducted to assess the effect of domain size on the turbulence statistics. We consider two cases: one from the very stable ($\Pi_w=6404,\Pi_s=0.25$) and a second one from the weakly stable regime ($\Pi_w=16200, \Pi_s=1$), see table \ref{tab:domain}. The former case is observed to be a critical parameter for $\Pi_w$ below which the flow remains in very stable regime irrespective of $\Pi_s$ value. On the other hand, the latter case shows continuous turbulence. The domain size is increased from 26D to 104D, by doubling the domain extent progressively. Note that the number of grid points are also doubled progressively to ensure that the simulations are fully resolved. 

\begin{table}
\footnotesize
\caption{Simulation parameters for investigating the effect of horizontal domain. From left: Case number, dominant stratification mechanism (DSM), $\Pi_w$, $\Pi_s$, Reynolds number, domain size, number of grid points in $x-y$, and $z$ directions, respectively, and surface Richardson number. The first three cases refers to the very stable (VS) regime, whereas the last three are in the weakly stable (WS) regime. } 
\label{tab:domain} 
\centering
\begin{tabular}{@{}lllccccc@{}}
\toprule
    Case no. & DSM & $\Pi_w$ & $\Pi_s$ & $Re_{D}$& Domain Size& $N_x^2 \times N_z$ & $Ri_s$  \\  \midrule        
VS-I \ & $N_a$ dom.  & 6404  & 0.25        & 400  &$26D \times 26D \times 24D$&   $128^2 \times 512$ & 0.0098 \\
VS-II \ & $N_a$ dom.  & 6404  & 0.25        & 400  &$52D \times 52D \times 24D$&   $256^2 \times 512$ & 0.0098 \\
VS-III \ & $N_a$ dom.  & 6404  & 0.25        & 400  &$104D \times 104D \times 24D$& $512^2 \times 512$ & 0.0098 \\ \hline
WS-I \ & $B_s \ \& \ N_a$        & 16200         & 1     & 400  &$26D \times 26D \times 24D$&   $128^2 \times 512$ & 0.0097 \\ 
WS-II \ & $B_s \ \& \ N_a$       & 16200         & 1      & 400  &$52D \times 52D \times 24D$&   $256^2 \times 512$ & 0.0097 \\  
WS-III \ & $B_s \ \& \ N_a$      & 16200         & 1 & 400  &$104D \times 104D \times 24D$& $512^2 \times 512$ & 0.0097 \\ \hline
\end{tabular}
\end{table}

\begin{figure}
    \centering
    \includegraphics[width=0.9\linewidth]{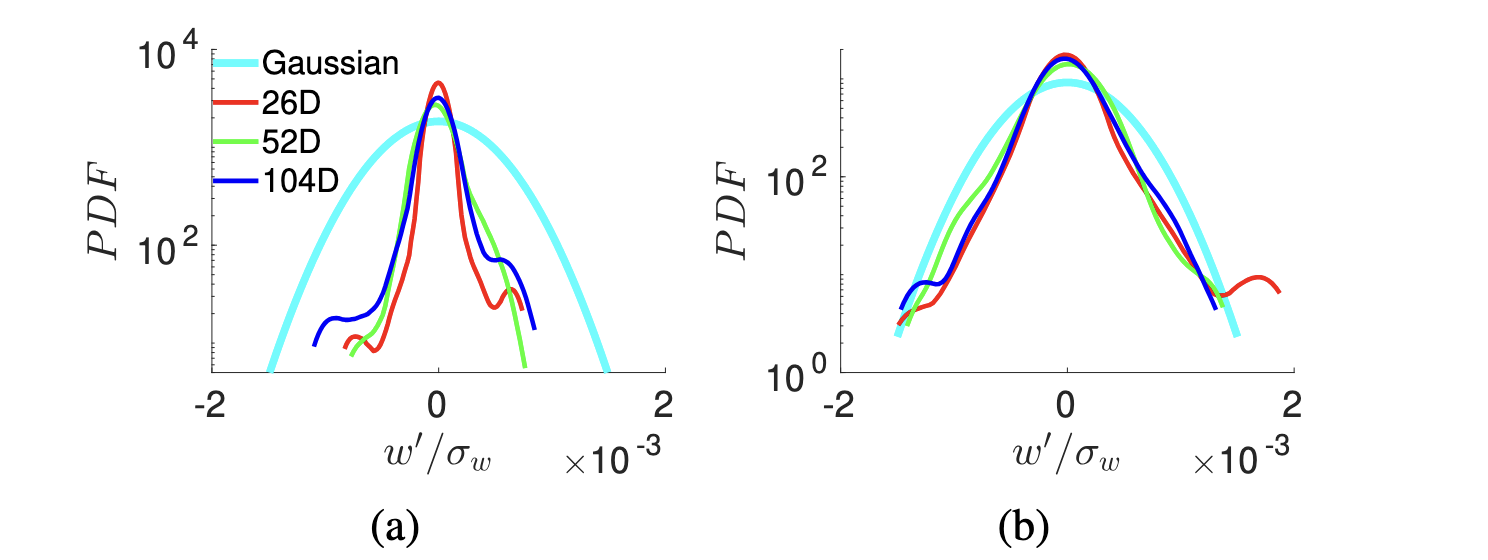}
        \caption{Effect of domain size on the PDF of vertical velocity fluctuations for (a) Case VS and (b) Case WS in table \ref{tab:domain}.  The two cases refer to the  very stable and weakly stable regimes. Analytical Gaussian distribution with zero mean and unit standard deviation is shown with cyan colored solid line.}
        \label{fig:PDF_domain}
\end{figure}

Figure \ref{fig:PDF_domain} shows PDF of vertical velocity fluctuation in the $x$–$z$ plane at $z/D = 0.5$ ($z^+ = 32$). Panel  (a) shows cases VS-I through VS-III, and (b) shows cases WS-I through WS-III. Table \ref{tab:domain} lists the parameters for each case.  For both the very stable (VS) and weakly stable (WS) cases, we observe that domain size does not influence the overall shape of the PDF and this qualitative aspect can be used to identify flow regimes. For instance, while PDF in WS cases shows wider distribution---signifying continuous turbulence---a narrow distribution in the very stable regime signifies turbulence collapse. Since the characterization of SABL into very stable and weakly stable regimes has been based on qualitative description of how turbulence manifest within the domain, we can use temporal statistics to identify flow regimes. Consequently, we rely on temporal statistics obtained from small domain $(26D \times 26D \times 24D)$ to identify the weakly and very stable regimes.

\section{Budget equations}\label{sec:appendix_budget}


We follow the presentation in \citet{shah_direct_2014} and reformulate the turbulent kinetic and potential energy budgets such that the presence of ambient stratification, $N_a$, appears in the budget equations. The TKE budget can be written as follows: 
\begin{eqnarray}
    \partial_t \langle q \rangle + \langle{u}_j\rangle \partial_j \langle q \rangle = &D_{q}& +~\epsilon_{q} + P_{q} + T_{q} + B_{q} + C_{q}+ \pi_{q} + \phi_{q}. \label{eq:tke_budget} 
\end{eqnarray}
Due to horizontal homogeneity, no subsidence, and $S^\prime_{ij}=0.5 (\partial_j u^\prime_i+ \partial_i u^\prime_j)$, the TKE budget terms on the right-hand side are written as follows: 
    \begin{eqnarray}
    B_{q} &=& g\langle{ w^\prime \Phi^\prime}\rangle, \hspace{4pt}  \\ 
    D_{q} &=& \nu \partial^2_{zz} \langle u^\prime_i u^\prime_i \rangle \hspace{4pt}, \hspace{4pt}  \\ 
    \epsilon_{q} &=& -\nu \langle \partial_j u^\prime_i \partial_j u^\prime_i \rangle  \hspace{4pt}, \hspace{4pt}  \\ 
    P_{q}  &=& -\langle u^\prime_i w^\prime\rangle \partial_z \langle u_i\rangle  \hspace{4pt}, \hspace{4pt}  \\ 
    T_{q} &=& -\frac{1}{2}~\partial_z \langle u^\prime_i u^\prime_i w^\prime \rangle \hspace{4pt},   \\ 
    C_{q} &=& -\frac{1}{2}~\epsilon_{ijz} \langle u^\prime_i u^\prime_j \rangle f, \hspace{4pt} \\ 
    \pi_{q} &=& -~\partial_z \langle p^\prime w^\prime \rangle, \hspace{4pt}\\ 
    \phi_{q} &=& 0, \hspace{4pt}
\end{eqnarray}

The turbulent potential energy (TPE) budget can be written as
\begin{eqnarray}
 \partial_t \langle q_{_\Phi} \rangle + \langle{u}_j\rangle \partial_j \langle q_{_\Phi} \rangle = &D_{_{\Phi}}& +~\epsilon_{_{\Phi}} + P_{_{\Phi}} + T_{_{\Phi}} + B_{\Phi}, \label{eq:pe_budget}
\end{eqnarray}
Similar to the TKE budget derivation, one can simplify the TPE budget equation under the horizontal homogeneity and no subsidence  conditions. The terms on the right-hand side simplify as follows:\
\begin{eqnarray}
B_{\Phi} = -\frac{N_a^2}{g}\langle{ w^\prime \Phi^\prime}\rangle, \label{eq:peb1} \\ 
\epsilon_{_{\Phi}}= -\beta \langle \partial_i \Phi^\prime \partial_i \Phi^\prime \rangle, \\ 
D_{_{\Phi}} = \beta \partial^2_{zz} \langle \Phi^{\prime~2} \rangle, \\ 
P_{_{\Phi}}  = -\langle w^\prime \Phi^\prime \rangle \partial_z \langle\Phi\rangle,  \label{eq:peb2} \\
T_{_{\Phi}} = -\frac{1}{2}~\partial_z \langle w^\prime \Phi^\prime \Phi^\prime \rangle  
\end{eqnarray}


\section{Linear stability analysis}\label{sec:appendix_lsa}
In this section, we describe the mathematical framework used for linear stability analysis and its validation. 
The non-linear system of equations (Eqs. \ref{eq:gov_eqn1}-\ref{eq:gov_eqn3}) can be linearized around a base flow state associated with a perturbation. We consider the perturbation equation of type $\phi=\hat{\phi}(z)e^{\text{i}(\kappa_x \cdot x +\kappa_y \cdot y)}e^{\omega t}$, where $\omega=\psi + \text{i} \sigma$ is a complex number whose real part signifies the temporal oscillation ($\psi$) and imaginary part shows the growth rate $(\sigma)$. For $U_g$, $D$, and $N^2_{a_c}=N_a^2+G_w$ as velocity, length and composite buoyancy scales, the non-dimensionalized form of linearized equations become:
{\large{
\begin{eqnarray}
\text{i} k \hat{u}  + \text{i} l \hat{v} + \frac{\partial \hat{w}}{\partial z}  &=&0 \\
\omega \hat{u} + \text{i} \hat{u} (k \bar{u}+l \bar{v})+\hat{w} \bar{u}_z &=& -\text{i} k \hat{p}+ \frac{1}{Re} \{-(k^2+l^2) + \partial^2_{zz} \} \hat{u} + \frac{2}{Re} \hat{v}\\ 
\omega \hat{v} + \text{i} \hat{v} (k \bar{u}+l \bar{v})+\hat{w} \bar{v}_z &=& -\text{i} l \hat{p}+ \frac{1}{Re} \{-(k^2+l^2) + \partial^2_{zz}\}\hat{v} - \frac{2}{Re} \hat{u}\\ 
\omega \hat{w} + \text{i} \hat{w} (k \bar{u}+l \bar{v}) &=& -\frac{\partial \hat{p}}{\partial z} + \frac{1}{Re} \{-(k^2+l^2) + \partial^2_{zz}\}\hat{w} + \frac{2(1+\Pi_s)}{\Pi_w \Pi_f}\hat{b} \\
\omega \hat{b} + \text{i} \hat{b} (k \bar{u}+ l \bar{v})+\hat{w}  &=& \frac{1}{RePr} \{-(k^2+l^2) + \partial^2_{zz}\} \hat{b} 
\end{eqnarray}
}}
where $Re=\sqrt{2\Pi_w/\Pi_f}$. 
The exact laminar Ekman solution and zero buoyancy is used as the base flow for velocities, and buoyancy, respectively. The base flow, marked with over-bar, is as follows:
{\large{
\begin{equation}
\bar{u} = 1-e^{-z} \cos{z}; \hspace{2pt}\bar{v} = e^{-z} \sin{z}; \hspace{2pt}\bar{w} = 0; \hspace{2pt} \text{and} \hspace{2pt}\bar{b}_z = 0
\end{equation}
}}

\begin{figure}
    \centering
    \includegraphics[width=0.6\linewidth]{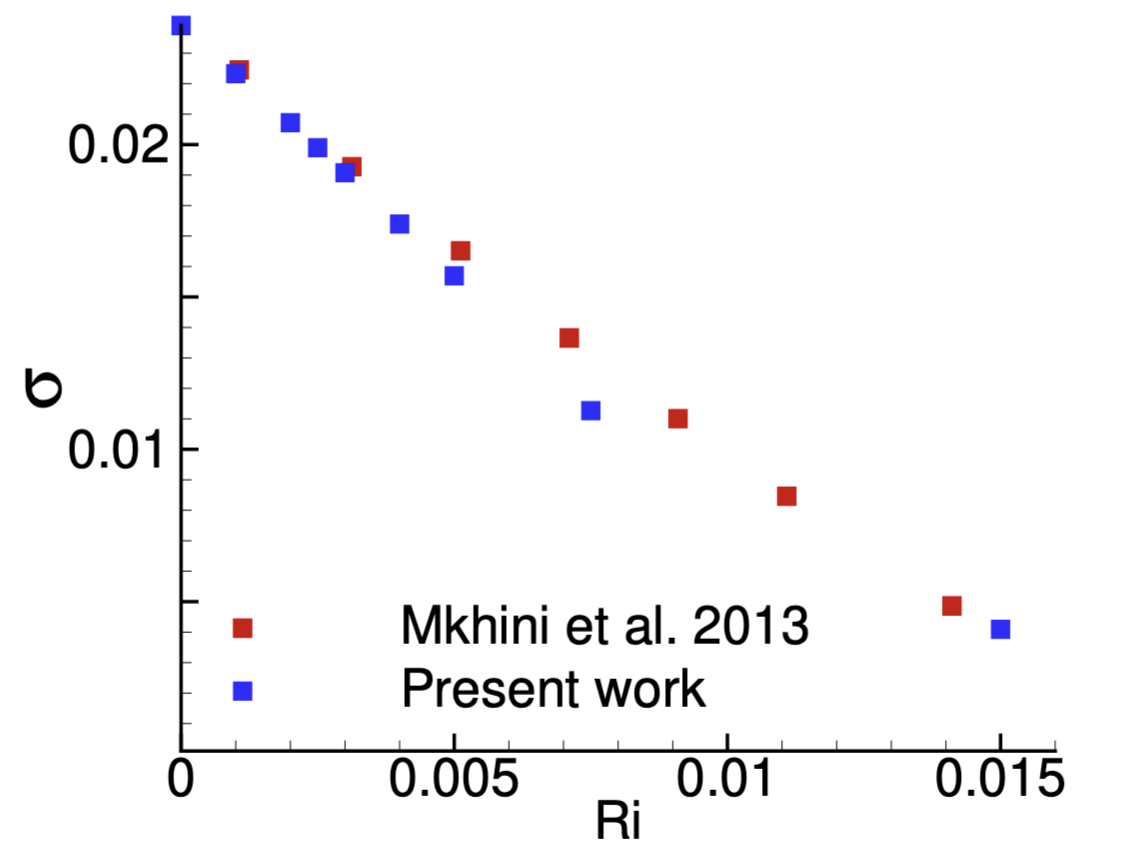}
    \caption{Comparison of growth rate computed from Nektar++ with \citet{mkhinini_secondary_2013} for stratified case. The slight deviation is attributed to the identification of local maximum growth.}
    \label{fig:lsa_validation}
\end{figure}

We have validated the LSA formulation with the work of \citep{mkhinini_secondary_2013}, see figure \ref{fig:lsa_validation}. Our data shows a good agreement, yet slight deviations can be attributed to the methodology employed by \citet{mkhinini_secondary_2013}, where they progressively identified maximum growth rate in the local neighbourhood of the previous $Ri$. In addition, we have also validated the numerical framework with primary instabilities  of neutral Ekman layer \citep{lilly_instability_1966, dubos_emergence_2008}.

\backsection[Funding]{This material is based upon work supported by the National Science Foundation under Grant No. (2203610) and by University of Pittsburgh Center for Research Computing and Data, RRID:SCR\_022735, through the resources provided. Specifically, this work used the H2P cluster, which is supported by NSF Award No. OAC-2117681.

This work also used Anvil at Purdue University through allocation EES240013 from the Advanced Cyberinfrastructure Coordination Ecosystem: Services \& Support (ACCESS) program, which is supported by U.S. National Science Foundation grants \#2138259, \#2138286, \#2138307, \#2137603, and \#2138296.}

\backsection[Declaration of AI use]
{During the preparation of this work the authors used Microsoft CoPilot and Claude (Sonnet 4.6) in order to assist with improving the clarity and quality of the English language. After using this tool/service, the authors reviewed and edited the content as needed and take full responsibility for the content of the publication.}

\backsection[Declaration of interests]{The authors report no conflict of interest.}

\bibliographystyle{jfm}
\bibliography{jfm,references}
\end{document}